# ANALYSIS OF SOLAR WIND ENERGY AND THE AKASOFU PARAMETER FOR ENERGY DYNAMICS ASSESSMENT DURING SUPERSUBSTORM

A dissertation

Submitted to the Dean Office, Institute of Science & Technology, Tribhuvan University, Kirtipur in the Partial Fulfillment for the Requirement of Master's Degree of Science in Physics.

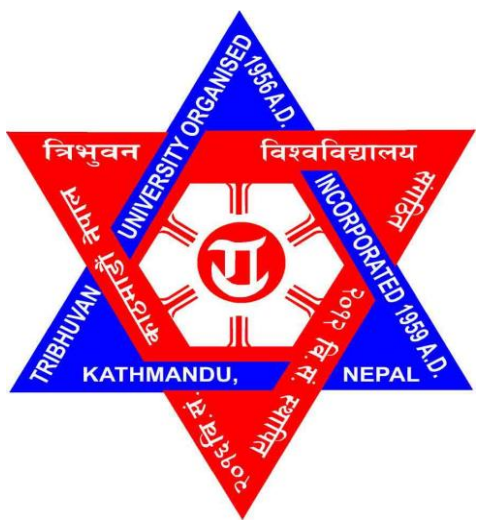

By
**Shailaja Wasti**

June, 2019

:

# Abstract

Solar activities have significant impact on the member of the solar systems including earth. As earth is a planet with its own magnetic field, solar emissions with magnetic fields do interact with the earth's own magnetic field to create geo-magnetic disturbances. Substorms are geo-magnetic disturbances of shorter duration (few hours or less) which are mainly concentrated in the auroral region and originates due to the ionospheric current injection at high latitudes. Supersubstorm are intense substorms as identified by peak activities in geo-magnetic indices. In this thesis, we have investigated the energy dynamics during the supersubstorm and analyzed the relation of the energy components with different geo-magnetic indices and interplanetary magnetic field (IMF) components. In particular, the solar wind energy levels and the energy coupled into the magnetosphere as captured by the Akasofu parameter are studied. A qualitative analysis of the computed energy and its relationship with different geo-magnetic indices is presented. Correlation and cross-correlation analysis have been used to get insights into the nature of the relationship between the energy components and indices. This for example uncovered the negative relationship between the solar wind energy and the Akasofu parameter. A summary analysis of the supersubstorm event in comparison with the quiet phases are made. This has led to insights for example showing the higher energy levels during supersubstorm and different energy ratios (Akasofu parameter divided by the solar wind energy) existing for different supersubstorm.

**Keywords:** *Supersubstorm, Akasofu, IMF, Solar wind energy*

**Contents**

# List of Figures

# List of Table

# Acronyms

| | |
|---|---|
| AE | Auroral Electrojet |
| CMEs | Coronal Mass Ejections |
| GSM | Geocentric Solar Magnetospheric |
| H | Horizontal |
| IMF | Interplanetary Magnetic Field |
| MHD | Magnetohydrodynamics |
| OMNI | Operating Mission as Nodes on Internet web system |
| PPs | Plasma Parcels |
| $R_E$ | Earth Radii |
| SC | Solar Cycle |
| SSC | Storm Sudden Commencement |
| SSS | Supersubstorm |
| SYM-H | Symmetric H-component |
| UT | Universal Time |

# Chapter 1 Introduction

## 1.1  Background

Solar activity has a major influence for the members of our solar system and beyond. Naturally, changes in solar activities bring about significant impact on earth. Sun can be basically described as a sphere of hot plasma mostly consisting of hydrogen and helium [1]. Nuclear fusion of hydrogen into helium is the energy generating process within the sun. Different regions can be demarcated in sun's body and its vicinity based upon the dominant activity occurring therein. For example, the interior of the sun is the solar core where the main energy is generated based on the nuclear fusion reaction. This energy is propagated further towards the sun's surface through radiation in a zone, aptly names as the radiative diffusion zone of the sun. Energy is release from the sun through different phenomenon like solar flares, coronal mass ejections (CMEs) etc. which releases energy which traverses through the interplanetary space. These energy releases also have magnetic content which can interact with the planet's own magnetic field, if the planet has one.

Earth has its own magnetic field and thus are subject to interactions with the magnetic property of the solar emissions, here primarily referred to as solar wind henceforth. In this work, we study the phenomenon of geomagnetic storm which are the magnetic disturbances brought by the solar emissions (CME, solar flares etc. travelling as solar wind) onto the earth's surface. The magnetic field activities in the earth are quantified using different geo-magnetic indices. Auroral Electrojet (AE) index is for example one such indices. AE index measures the magnetic activity in the auroral zone due to the current in the ionosphere around the auroral region [2]. The current in the auroral zone is changed with the incoming solar wind. Therefore, AE index measures, in a way, the impact of the incoming solar wind emissions. In the technical terms, AE index is computed as the range (highest – lowest) of the registered magnetic activity (measured by the horizontal magnetic field). A geomagnetic storm will be registered as aberrations also in the AE index. In this work, we study supersubstorm which is characterized by very high activity levels in the geomagnetic indices. Though other indices can also be used to identify/denote supersubstorm, in this work we define supersubstorm based on the registered AE index levels. Supersubstorm, as the name suggests, are intense substorms. In terms of physical interpretations, substorms are differentiated from a geomagnetic storm based on the duration (few hours of substorm compared to several days of a magnetic storm) and the concentration regions (substorm are mainly concentrated in the polar region and observable there, while geomagnetic storms are observable everywhere) [3].

Supersubstorm are interesting to study not only because of the need to understand the solar magnetohydronamics and the physics of its emissions, but also because of the significant practical impact supersubstorms bring into earth. Supersubstorm could have a major impact on the electromagnetic operations in the earth, radio communications, space communications, weather monitoring applications etc. In this work, we analyze the energy dynamics of a supersubstorm. We study the total energy content of the solar wind and the energy that is coupled into the magnetosphere as measured by the Akasofu parameter (epsilon) which has been described in [4]. The energy dynamics are studied in relation to different interplanetary magnetic field (IMF) parameters and geo-magnetic indices. We take four event days, three comprising supersubstorm and one quiet day, and study different energy and geomagnetic index behaviors on these days. A qualitative analysis of the observed energy and the indices on different days, with and without supersubstorm, gives a fine-grained look into the phenomenon of supersubstorm (e.g. how the energy builds up and leads to changes in geo-magnetic indices). This qualitative analysis is further extended with correlation analysis between different variables (energy and geo-magnetic indices). Correlation analysis helps uncover the dependency relation between energy components and geo-magnetic indices. For example, how strong are changes in AE-index when there are changes in the solar energy can be understood by observing the correlation between these variables.

As correlation analysis can only uncover direct relationship, we also use cross-correlation relationship to investigate the presence of time-lagged relationship between energy components and geo-magnetic indices. This analysis sheds light on the time lag between changes in energy level and the time when these changes are reflected as changes in the geo-magnetic indices, e.g as investigated in [5]. The results from the cross-correlation analysis are presented in this work in context of further studying the energy dynamics. We also report the summary statistics representing properties of a supersubstorm event (e.g. average energy levels) and put it in context of results from a quiet day in order to produce global characterization of a supersubstorm event.

This thesis is organized as follows. A brief literature review on solar activities, geomagnetic storm and supersubstorm is presented in Section 1.2. The theoretical background of the work describing various physical phenomenon is presented in Chapter 2. There we briefly describe the sun and solar phenomenon in relation to the supersubstorm activity before describing the geo-magnetic indices and supersubstorm phenomena. The dataset, methods used for energy computation and data processing used for the analysis in this work are presented in Chapter 3. The results obtained are presented in Chapter 4 and the conclusions along with the future work are presented in Chapter 5.

The focus of our research is achieving the following main objectives:
- Analysis of the solar wind energy for energy dynamics assessment during supersubstrom.
- Calculation of solar wind energy using Akasofu's parameter during supersubstorm.

## 1.2 Literature review

The solar wind is the ultimate source of energy that drives the magnetospheric dynamics. It is necessary to understand the complicated current system in the magnetosphere and ionosphere because the solar wind has huge amount of charged particle and those charged particles interact to Earth magnetic field affect life on the Earth. Numerous studies have been carried out in the past decades regarding the phenomenon on solar wind-magnetosphere interactions. Chapman et.al., [6] studied the formation of magnetosphere where solar wind is considered as an unmagnetized plasma which simply flows around the earth. Dungey [7] studied about the open magnetosphere for the first time where the magnetic flux from the magnetosphere is interconnected with the interplanetary magnetic flux. This model provides correctly how solar wind couples its energy to the magnetosphere by a dynamo process. A mechanism on energy flux of an enhanced solar wind was later reviewed by Parker [8] in which the heating of magnetospheric plasma by hydromagnetic waves related with solar wind was studied.

In the study on the development of the main phase of magnetic storms by Akasofu and Chapman [9] for number of storms, it is concluded that the development of the storms varies between the solar streams far beyond from a mere difference between their pressures. In the following paper Akasofu [10] studied about the magnetic storm and finds that the development of main phase is associated with intense polar magnetic disturbances and suggests that the ring current is formed during intense magnetospheric substorms.

Hewis and Bravo [11] mapped 96 heliospheric disturbances during august 1978-september 1979 to analyze the sources of disturbances. In this study it was found that the erupting streams generated by coronal hole activity were accomplished by shocks and geomagnetic activity and the geomagnetic storms are more associated coronal holes than the solar flares. Following this Burlaga [12] studied the geomagnetic disturbances during the period 1968-1986 and found that the largest geomagnetic storm was occurred on july 13, 1982 and between 1972 and 1983 there occurred 30 geomagnetic storms with Ap > 90. These results suggest that geomagnetic disturbances are associated with compound streams or magnetic clouds.

Fast CMEs are the most dominant interplanetary phenomenon associated with very intense magnetic storms (-250nT ≤ Dst < -100nT). Gonzalez [13] states that the structures for intense magnetic storms of the IMF are the sheath region behind the fast forward interplanetary shock and CME ejected itself. Not only this but other aspect that contribute to the intense magnetic storms are the magnetic clouds with very intense core magnetic fields that has a substantial southward component [14] which tends to have large velocities. Both the field and the velocity are responsible for large amplitude interplanetary dawn-dusk electric fields which at extreme conditions leads to very intense magnetic storms. Later in a research by Rawat [15] found that a prominent scatter occurs when the magnetospheric electric field is directed from dusk to down which suggests that for the interplanetary conditions in the solar wind low-latitude ground magnetic data serve as a proxy. In the same study it is concluded that intensity of magnetic storms is controlled by duration and magnitude of the southward IMF $B_z$ and large energy input with long duration southward $B_z$ leads to the development of super intense magnetic storms (Dst ≤ -200nT).

Richardson [16] studied the solar wind structures driving geomagnetic storms over nearly three solar cycles (1972 -2000) and found that almost 97% of the intense storms are generated by transient structures associated with CMEs. While in case of weaker storms, solar minimum is associated with streams and solar minimum is with CMEs which reflect the change in the structure of the solar wind. Energy coupling between solar wind and magnetosphere is believe to depend only on solar wind speed and magnetic field but recent studies indicate that density has also an important role in solar wind-magnetosphere interaction. Lopez et al., [17] use the global magneto-hydro-dynamics (MHD) simulations of the solar wind magnetosphere interaction and found that during strongly southward IMF even when solar wind magnetic is steady higher densities result in a larger compression ratio across the shock that produces larger magnetosheath fields.

The relationship between solar wind conditions and geomagnetic disturbances have been studied from several decades. Xu et al., [18] studied the correlation between solar wind and geomagnetic storm from 1998 to 2008 and found that during solar maximum interplanetary coronal mass ejections (ICMEs) primarily magnetic clouds are the dominant feature while corotating interaction regions (CIRs) is during solar minimum. Solar wind became unstable during rising and declining phase in which ICMEs mainly centered on the maximum period that occurred 1.3 times of that of CIRs.

Among different geomagnetic indices auroral electrojet (AE) index is the crucial physical parameters in monitoring and predicting magnetospheric substorms. Ziming et al., [19] use the AE index in 2004 and found its relation with substorms. This study shows three kinds of AE events: simultaneous rapid increase of westward and eastward AE, rapid increase of westward and unchangeable eastward AE, and rapid increase of eastward and unchangeable westward AE. This research concluded that most of AE events correspond to substorms but only a few are not accompanied by substorms.

Tsurutani et al., [20] analyzed supersubstorm (SSS) events for solar cycle (SC) 23 (1996 - 2009) and found that 57% of the SSS events were associated with superstorms, 40% with intense storm and 3% with non-storm ( Dst ≥ -50nT). Among 11 superstorms during SC 23, 9 (~82%) are associated with SSS events which shows that there is some relation between SSS and superstorms but there is no one to one correspondence between these two phenomena. This result shows that SSS events are externally triggered by very high intensity (~ 30 to 50 cm-3) solar wind plasma parcels (PP) heating the magnetosphere. After the analysis of the results from SC 23 Tsurutani et al., [20] put forward an hypothesis that southward fields input energy into the magnetosphere/ magnetotail and PPs trigger the release of stored energy. Later, Hajra et al., [21] studied magnetic storm of approximately 3 SC (1981 - 2012) and found that 77% of the SSSs were associated with small region of PPs with very high density impinging upon magnetosphere. This analysis concluded that there is no strong relationship between the occurrence and strength of SSS events with the intensity of magnetic storm.

# Chapter 2 Theoretical background

## 2.1 The sun

The sun is a sphere of hot plasma which is compressed by its own gravitation attraction. It is mainly divided into three regions: Sun's interior, solar atmosphere, and the visible "surface" of the sun. Sun's interior consists of three regions namely the core, the radiative zone, the convective zone. Core is the central part where nuclear fusion reactions occur. The radiative zone is the middle part from which energy is carried outward through this layer, carried by photons as thermal radiations. The convection zones are the outer layer of the sun from which energy is transported to the solar surface. A strong shear layer exists at the boundary of the convection zone [22] in which sun's magnetic field is generated by a dynamo. The solar atmosphere from which photons can escape directly into space comprised of three regions namely photosphere, chromosphere, and corona. Between the Sun's interior and the solar atmosphere there is the Sun's thin visible surface layer of plasma known as the photosphere. Photosphere is dense and opaque from which most of the solar radiation emits. The lower region of the solar atmosphere consists of the chromosphere whereas the upper region consists of the corona. Corona is hotter than the photosphere which gradually turns into the solar wind [23]. The entire structure of the solar interior and atmosphere indicating the sizes of the various regions, temperatures and densities is shown in figure 2.1.

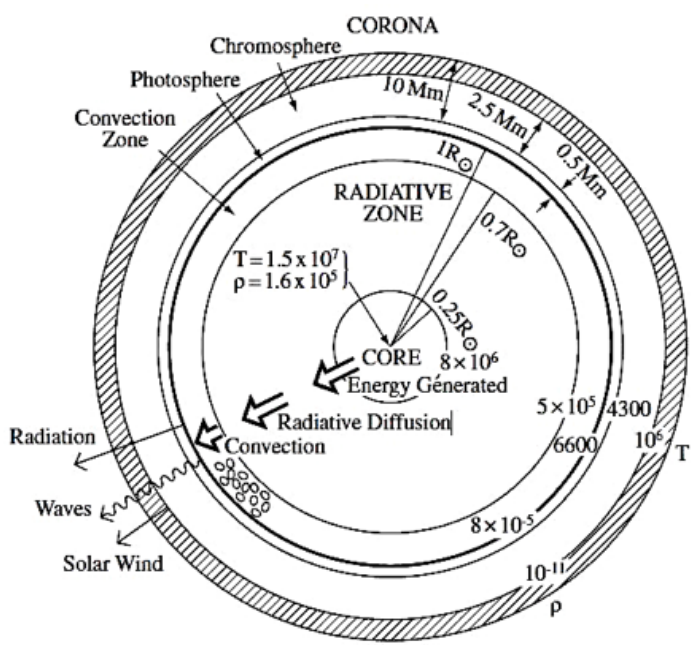

**Figure 2.1** Overall structure of solar interior and atmosphere [23]

## 2.2 Solar wind

The corona is continuously expanding outward as a stream of ionized particles called the solar wind. Solar wind comprises of neutral components that penetrates across the magnetopause and becomes energetic protons after exchanging charge with $H^+$ and $O^+$ ions in the thermosphere [24]. The solar wind is divided into two components i.e., the slow solar wind (300 to 400 kms$^{-1}$) related to the streamer belt region and the fast solar wind (700 to 750 kms$^{-1}$) originating from coronal holes [25]. Slow solar wind which has high density and low temperature is highly variable that comes from streamer belt at the time of solar minimum and small coronal holes and active regions at the time of solar maximum. Fast solar wind of lower density and higher temperature is relatively steady and occupies a polar region at solar minimum and occurs at all latitudes at solar maximum. This fast solar wind in polar holes accelerates rapidly and reach half of its terminal speed by heights of 2 to 4 $R_\odot$ [26]. The solar wind is the ultimate source of energy for all the magnetospheric dynamics [7]. The stream of solar wind and the electric currents inside the magnetosphere determine the shape of magnetosphere [27].

## 2.3 Solar activity

The sun is in variable magnetic field that continually well up into the solar atmosphere as a result of dynamo action in the solar interior [28, 29]. Solar activity arises from magnetic field through the different layer of solar atmosphere and into interplanetary space. The phenomenon that occurred in the solar surface in the form of solar flares, sunspots, prominences, coronal mass ejection (CME) is known as solar activity. One solar cycle spans ∼ 11 years on average but this duration is not exactly 11 years, it has been sometimes being short cycle of 9.5 years and sometimes as long as 12.5 years [30]. Changes occurs in the Sun's internal magnetic field and surface disturbances level during a solar cycle. On the basis of maximum and minimum number of sunspots on the sun's photosphere solar cycle is named as solar maximum and solar minimum respectively.

### 2.3.1 Solar minima

The solar minimum is the period of least solar activity in the suns 11 years of solar cycle. During this period occurrence of sunspots, coronal mass ejections and solar flares decreases in frequency, sometimes not occurring for days on end.

### 2.3.2 Solar maxima

The solar maximum is the period of greatest solar activity in the 11 years of solar cycle. During this period, major fluctuations in sun's magnetic field, and an increase in the number of sunspots, coronal mass ejections, and solar flares can be observed. Coronal mass ejections and emission of high energy solar flares become more intense during the period of solar maximum.

### 2.3.3 Solar flare

A solar flare is a sudden release of magnetic energy built in the solar atmosphere in a rapid and intense variation. Flares produce a burst of radiation across the electromagnetic spectrum and emits all wavelengths from radio waves to X-rays and gamma-rays. The occurrence of solar flares is high during solar active period and low during low period.

### 2.3.4 Coronal mass ejection

CME are the powerful eruption of plasma and magnetic field from the solar corona which inject large amounts of mass and energy into interplanetary medium. CMEs which is identified as upward mass motion from solar corona was discovered in 1971 [31] using the seventh orbiting solar observatory (OSO-7) that emits mass and energy of the order $\sim 10^{16}$ g and $10^{32}$ ergs [32]. The largest and most energetic eruption that cause disturbances in the flow of solar wind create geomagnetic storms. Rearrangement takes place between two oppositely charged particles and after this rearrangement the energy stored in the oppositely directed magnetic field lines is released which is responsible for solar flares that drives the CME. Many researchers studied about this sudden energy release that are associated with solar flares and CME [33, 34]. CME are observed at all latitudes which occurs both inside and outside of the active regions. When CME occurs outside the active regions magnetic field is weak whereas inside the active regions magnetic field is strong with higher plasma and particle velocities. The CME which tracks the sunspot number is high during the duration of solar maximum [35]. At the active regions of the sun from where solar flares and CMEs are originated the magnetic field is stronger.

## 2.4 Magnetosphere

The magnetosphere is the region of space surrounding earth whose magnetic field is the magnetic field of the earth rather than the magnetic field of the interplanetary space. Earth surface temperature is determined by the radiant energy from the sun which supplies the energy for natural process on the Earth's surface and its atmosphere [36]. Upper atmosphere of the earth is ionized by radiations from the sun where electromagnetic radiations travel directly between the Sun and earth at the free space speed with a transient time of approximately 8.3 minutes. Along with the electromagnetic radiations stream of charged corpuscles is also emitted from the sun which is called the solar wind. The magnetized collisionless solar wind interacts with the Earth's magnetic field to a region around the planet to forms a magnetosphere. The input from interplanetary space to the magnetosphere differs significantly as it depends on where reconnection happens and also depends on how it reconnects magnetic flux from interplanetary space [37].

## 2.5 Magnetic reconnection

The solar wind that carry magnetic fields interacts with a planetary magnetic field with different velocity. The process that allows solar wind plasma to enter the magnetopause and then to the magnetosphere is merging between solar wind and the terrestrial magnetic fields which is known as magnetic reconnection [38]. Magnetic energy is converted into plasma kinetic energy which is redistributed within the various region of magnetosphere during magnetic reconnection process.

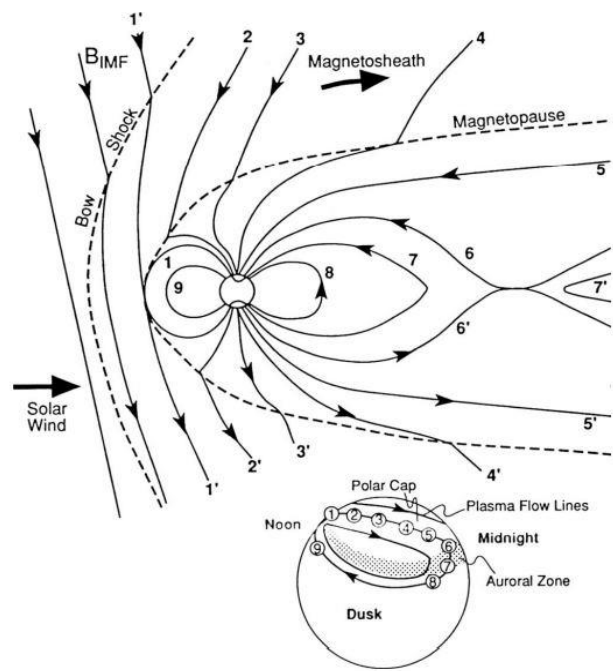

**Figure 2.2** Magnetic reconnection and generated flows of plasma in the magnetosphere [39]

The boundary layer that separates the geomagnetic field from the solar wind plasma is the magnetopause. Pressure balance is maintained between incoming solar wind and the earth's magnetic field on the dayside magnetopause while on the nightside lines of force stretch behind the Earth in a direction away from the sun. In figure 2.2 it is seen that IMF as indicated by field line 1' reconnects the Earth dipole indicated by field line 1 at the magnetopause. Also, in the same figure field line 6 and 6' reconnects in the tail.

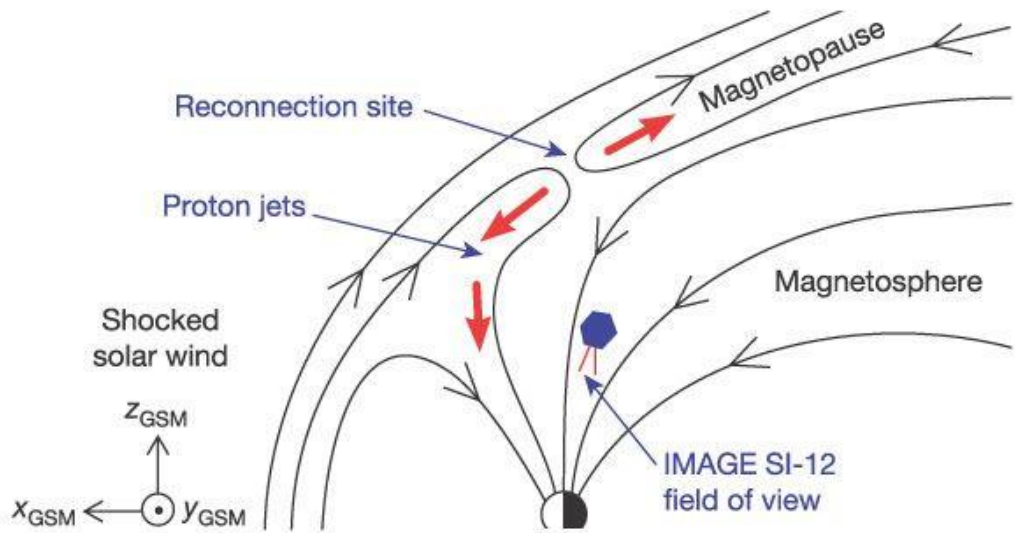

**Figure 2.3** Magnetic reconnection at the magnetopause and its footprint in the ionosphere [40]

Findings on the dayside proton auroral spot [41, 42], which is created by solar wind protons leaking, is accelerated to the magnetopause through reconnection. In figure 2.3, Magnetic reconnection occurred between solar wind magnetic field along z axis of the geocentric solar magnetospheric (GSM) coordinate system and the opposite magnetospheric field lines.

## 2.6 Geomagnetic disturbances

### 2.6.1 Geomagnetic storm

Geomagnetic storm is the temporary disturbance in the geomagnetic field after the impact of the solar emissions on the earth's magnetosphere. Geomagnetic storms are mainly caused by solar wind transients from CMEs, solar flares, high speed streams (corotating interaction regions) and magnetic clouds [43] during the interaction between high and low speed streams. Solar wind energy is transferred into magnetosphere to produced geomagnetic storm by magnetic reconnection process between southwardly oriented IMF Bz component and the opposite magnetospheric field lines [4, 7]. The development of geomagnetic storm is divided into three phases i.e., initial phase, main phase, and the recovery phase. The initial phase caused by the increased solar wind ram pressure usually occurs after the storm sudden commencement (SSC) [44] where the horizontal (H) components in low latitudes increases discontinuously and last for few hours. Initial phase is followed by main phase where H components decrease with greater magnitude than the initial phase and last for hours to days. During the main phase when the southward component of the IMF causes magnetic reconnection at the magnetopause there is an increase in the ring current injection with energetic solar wind plasma. Main phase is followed by the recovery phase where the H components recovers slowly reducing the injection of particles into the ring current until it attains the minimum value.

### 2.6.2 Geomagnetic substorm

Geomagnetic substorm is a disturbance in the earth's magnetosphere originated on the nightside of the earth where energy derived from solar wind-magnetosphere interaction is deposited into the earth's magnetosphere and ionosphere [24, 45, 46]. Substorms are the integral part of magnetic storm [47] however substorm can occur independently of storms [48].

### 2.6.3 Geomagnetic supersubstorm

Geomagnetic supersubstorm are intense substorms characterized by very high activity levels in the geo-magnetic indices. Generally, the SuperMAG AL (SML) index activity is used to characterize the occurrence of supersubstorm. For example, a criterion of SML < -2500 nT is used to identify supersubstorm [49]. Similar activity levels can also be defined for the AE index (Auroral Electroject index), as has been used in the works [50].

In this work we consider the AE > 2500 nT as the supersubstorm phase though other works have considered even a lower level to fall into the supersubstorm category, e.g. Gregorczyk [50] has taken AE > 1500 nT as supersubstorm. Also in Bruce et.al., [51] the authors suggest superstorm to be occurring when AE > 1500 nT. However in Gjerloev [52] authors have shown that AE index are only slightly smaller than the SML index. Therefore, the commonly used level of 2500 nT might as well be used for the identification of the supersubstorm using AE index alone. The advantage of keeping the highest levels in the AE index for the demarcation of the supersubstorm event is that the selected event phase would definitely be a supersubstorm as per any of the threshold category that has been used in the literature.

# Chapter 3 Methods and Data

## 3.1 Datasets

The dataset required for the analysis and study in this work was obtained using the freely accessible OMNI website (http://www.omni-web.gsfc.nasa.gov/). Using the web-interface we downloaded several IMF (interplanetary magnetic field) parameters and geo-magnetic indices as relevant for this work. The data were obtained on a 1-minute sampling interval. Therefore, our analysis produces result per minute.

The dataset that has been made available in the OMNI website has been sourced from multiple spacecraft and earth stations. The details about these data can be obtained from the website of OMNI. We extract data files for selected time consisting of 16 main variables. These variables are:

- Year: year when the data was collected
- Day: Day of the year for the data collection
- Hour
- Minute
- Field magnitude average (in nT)
- BX (nT): X component of the magnetic field
- BY (nT): Y component of the magnetic field
- BZ (nT): Z component of the magnetic field
- Speed (km/s): solar wind speed
- Proton density (in n/cc)
- Temperature (in kelvin)
- Flow pressure (nPa)
- Electric field
- AE-index (nT)
- AU-index (nT)
- SYM/H (nT)

Though we have downloaded a number of geomagnetic indices, not all of them will be considered for the analysis. This is further described in Section 3.3. It is to be noted that the data as obtained using the OMNI website has all the required time corrections to account for the propagation time between the spacecraft (where the measurement was taken) to the bow shock.

## 3.2 Akasofu's parameter and solar wind energy

In this study we used coupling function ε(W) as formulated by [53] to estimate solar wind energy. The purpose of using ε(W) in our research is that this function considers the reconnection process. The main phenomenon for energy transfer from solar wind to the magnetosphere is the magnetic reconnection process. The empirical formulation of the coupling function and total energy input by [53] is

$$\varepsilon = 10^7 u B^2 l_o^2 sin^4\left(\frac{\theta}{2}\right) \; [W] \quad (3.1)$$

$$W_\varepsilon = \int_{t_0}^{t_m} \varepsilon dt \; [J] \quad (3.2)$$

Where, B is IMF strength, θ is the IMF clock angle in the plane perpendicular to Sun-Earth line, $l_o$ is empirically determined factor i.e., $l_0 = 7R_E$, $R_E$ is Earth radii.

IMF clock angle (θ) is determined as

$$\theta = \tan^{-1}\left(\frac{B_y}{B_z}\right) \quad (3.3)$$

The total energy input is obtained by integrating ε over the main phase from $t_0$ to $t_m$ of each magnetic supersubstorm.

The energy produced by solar wind-magnetosphere interaction is dissipated into the different regions of the magnetosphere. This energy is partially deposited in the inner magnetosphere and some part in the auroral ionosphere. Energy in the auroral ionosphere is deposited as heat energy which arises partly from Joule heating and partly from the impact of auroral particle [4]. In equation (1) it is considered that the ε is just the total power input into the magnetosphere. Also, in case of radius of the dayside magnetopause ($l_o$), it does not have a clear meaning. Considering these all [54] modified equation (1) by replacing the constant value of $l_o$. [54] took into account that the radius of the dayside magnetopause depends on the solar wind ram pressure.

The energy produced by solar wind-magnetosphere interaction is then estimated by the corrected form of ε parameter based on the solar wind ram pressure i.e.,

$$\varepsilon^* = \left(\frac{R_{CF}}{l_o}\right)^2 \varepsilon \quad (3.4)$$

Where, R$_{CF}$, obtained from a balance between the kinetic plasma pressure and the magnetic pressure, is estimated as

$$R_{CF} = \left(\frac{B_o^2}{4\pi\rho v^2}\right)^{\frac{1}{6}} R_E \qquad (3.5)$$

Where, B$_o$ ~ 0.3 gauss, ρ is the mass density of solar wind, $v$ is the velocity of solar wind, R$_E$ is the radius of the Earth.

## 3.3 Geomagnetic indices

Geomagnetic indices are parameters which indicate the variations in the geomagnetic field. A number of indices are used in the literature in order to study various events of geomagnetic relevance. In this work, the main indices which are relevant are:

- AE index
- SYM-H index

Auroral Electrojet (AE) index is computed as the range (highest – lowest) of the registered magnetic activity (measured by the horizontal magnetic field). AE index measures the magnetic activity in the auroral zone. This measured magnetic activity is originated mainly due to the current in the ionosphere which are concentrated in the auroral region.

Symmetric H-component (SYM-H) index indicates the total ring current intensity during geomagnetic storms. The measurement for SYM-H is derived from the measurements as obtained using the ground-based magnetometers.

Besides the geomagnetic indices, we also use multiple IMF parameters (solar wind speed, IMF field components etc.) as has been described in Section 3.1.

## 3.4 Data processing

The data that has been downloaded from the OMNI website are available in the lst format. Lst format is mainly a text file format with data in a list-like form. Therefore, we parse it as a regular text file and read out the value of different variables which as available in different columns. Along with the data, the corresponding column information are also available which are downloaded as the fmt format. This consists of the name of the variables which are available in the lst file. By reading the fmt file and the lst file, all the data can be parsed and identified as being for a particular variable. We download data for days of particular interest. As the data are available with the sampling period of 1-minute, each day's recording consists of 1440 rows (observations).

After the data is read from the lst file, these are already accessible for direct analysis. However, there are data gaps in the recordings. Whenever there is a missing observation for a variable in any minute, this is stored as value -9999. We process the data to identify all the location of -9999 values and fill these values with the interpolated data. The interpolation is done based on a linear fit of the time variable and the observed variable values for the variable of the interest. E.g. if there are some observations in the AE-index column as -9999, first these locations are identified. A regression is fit from the remaining data by providing the linear time-index as the independent variable and the AE-index as the dependent variable. Once the linear regression fit is obtained, the time-index of the missing data point is fed into this regression line in order to obtain the interpolated values to fill the missing points.

No other further processing like smoothing and filtering of the variable is done.

## 3.5    Correlation Analysis

One of the statistical analysis we do in this work to study the relation between different variables is the correlation analysis. The correlation is obtained using the pearson correlation coefficient which is obtained as:

$$r = \frac{1}{n-1}\sum \frac{(x_i - \bar{X})(y_i - \bar{Y})}{s_x s_y} \qquad (3.6)$$

Where r captures the correlation between two variables x and y each containing 'n' observations in total. Here Sx and Sy indicates the standard deviation of the variables. The value of the pearson's correlation coefficient ranges from -1 to +1. The value of -1 for correlation coefficient indicates a correlated but inverted relation between the variables. The value of +1 indicates a relation where both variable changes together in the same direction. A correlation coefficient of 0 indicates no relation between the variables.

## 3.6 Event selection

For the analysis considered in this work, we selected 4 event days. These are listed below.

**Table 3.1** List of selected events

| Events  | Date              | Type          |
|---------|-------------------|---------------|
| Event-1 | 24 November, 2001 | Supersubstorm |
| Event-2 | 21 January, 2005  | Supersubstorm |
| Event-3 | 24 August, 2005   | Supersubstorm |
| Event-4 | 19 July, 2006     | Quite         |

There are three event days where supersubstorm activity was seen. There is one event day which is identified as a quiet day. The data from these events are helpful to study the supersubstorm energy dynamics in light with the dynamics as seen on a quiet day.

# Chapter 4 Result and Discussion

## 4.1 Qualitative analysis of the energy parameters and the geomagnetic indices

The plots of various IMF parameters, geomagnetic indices, and the correspondingly calculated values of the solar wind energy and the coupling parameter (epsilon) are first presented in this section. We present a qualitative and quantitative analysis of the different IMF and energy parameters as observed during different supersubstorm events.

**Event-1 Qualitative Analysis:  Supersubstorm on 24 November 2001.**

The IMF parameters, geo-magnetic indices, and the computed energy for this day is shown in Figure 4.1. Overall, there seems to be a pretty stable phase during the start of the day up until around 05:35 UT when for example a rise in solar wind speed (see the first panel) is seen. At around this same time, a change in activity as reflected by the different geomagnetic indices (Bz in the 2$^{nd}$ panel, Sym-H in the 6$^{th}$ panel and AE index in the 7$^{th}$ panel), electric field (Ey in the 3$^{rd}$ panel) and thus the dependent computed energy (epsilon in the 4$^{th}$ panel and solar wind energy in the 5$^{th}$ panel) are seen. The main variations seen in the activity subsides towards the end of the day, mostly after around 16:40 UT. Before 05:35 UT, the solar wind speed is in the range of 450 km/s. This rises to the level of around 900 km/s after the said time and reaches to the maximum of 1059.0 km/s at 15:15 UT. The average solar wind speed during this event is 775.11 km/s. As the rise in solar wind speed is seen, a corresponding decrease in the southward component of the magnetic field (Bz) is seen (2$^{nd}$ panel). The value of Bz which is around 0 nT in the start of the day falls to a low of -42 nT at around 07:06 UT. Then again a rise in the value of Bz is seen which grows to a high of 56 nT at around 10:26 UT before falling again to a similar lows of -40 nT. The value of Bz then recovers over the day settling to an average value of 10 nT. The maximum and minimum of Bz are registered as 64.11 nT and -45.43 nT which occur at 09:48 UT and 11:49 UT respectively. The electric field (Ey) shows an almost mirror-like activity in comparison of Bz (3$^{rd}$ panel), increasing when Bz decreases and vice versa. The maximum of the electric field is seen upto the value of 39.23 mV/m at 11:49 UT and the minimum is -57.85 mV/m seen at 09:48 UT.

The geomagnetic index SYM-H, the symmetric component of the ring current, is plotted in the 6$^{th}$ panel. The value of SYM-H is on average around -25 nT before rising slightly and falling significantly after around 05:50 UT. The maximum of SYM-H is reached to a value of 74 nT at 06:04 UT while the decrease in the SYM-H goes as low as -234 nT at 12:37 UT. The AE index during the event is shown in the 7$^{th}$ panel. Overall, there seems to be two main peak activity in the AE-index, the first starting around 05:56 UT and the second starting around 13:45 UT. In the first peak instant, the highest value of AE index reaches to 3543 nT at 07:12 UT. In the second peak instant, the AE index reaches upto 3230 nT at 13:52 UT.

The solar wind energy computed for Event-1 is shown in the 4$^{th}$ panel. When the increase of solar wind speed is seen around 05:35, correspondingly a higher value in the solar wind energy is seen. The energy reaches a highest value of 0.71 * 10^14 Joules/sec at 07:10 UT. The value of energy increases and then slowly subsides over the day to the low level of around 0.01 * 10^14 Joules/sec towards the end. The behavior of the epsilon (shown in the 4$^{th}$ panel) is a bit different with two distinct phases of high value one starting at around 05:39 UT and the other starting at around 10:37 UT. In the first phase, the value of epsilon reaches to the high of 0.451 * 10^14 joules/sec and in the second phase it reaches to the high value of 0.55 * 10^14 joules/sec.

The supersubstorm activity during this day according to the definition based on the AE index is shown in the Figure 4.2. In the plot, the shaded region corresponds to the supersubstorm activity region according to the used definition based on AE activity. Barring minor aberrations, there are two main supersubstorm activity during this day. The first event starts at 07:02 UT and ends at 07:15 UT. The duration of the first supersubstorm event is therefore 13 minutes. The second event starts at 13:47 UT and ends at 14:03 UT. This corresponds to a supersubstorm event duration of 16 minutes.

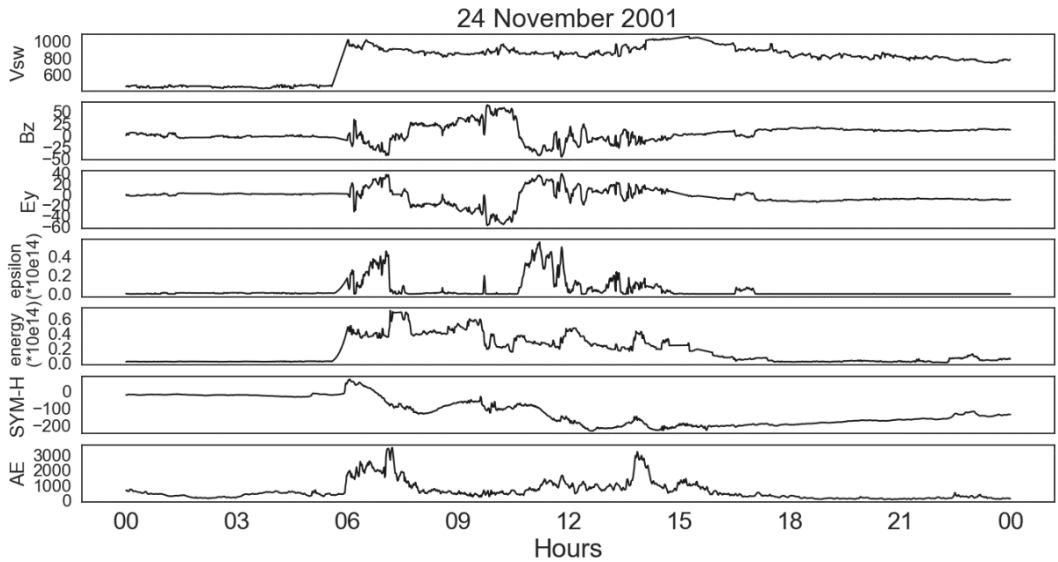

**Figure 4.1** IMF parameters and energy computed for 24 November 2001 data

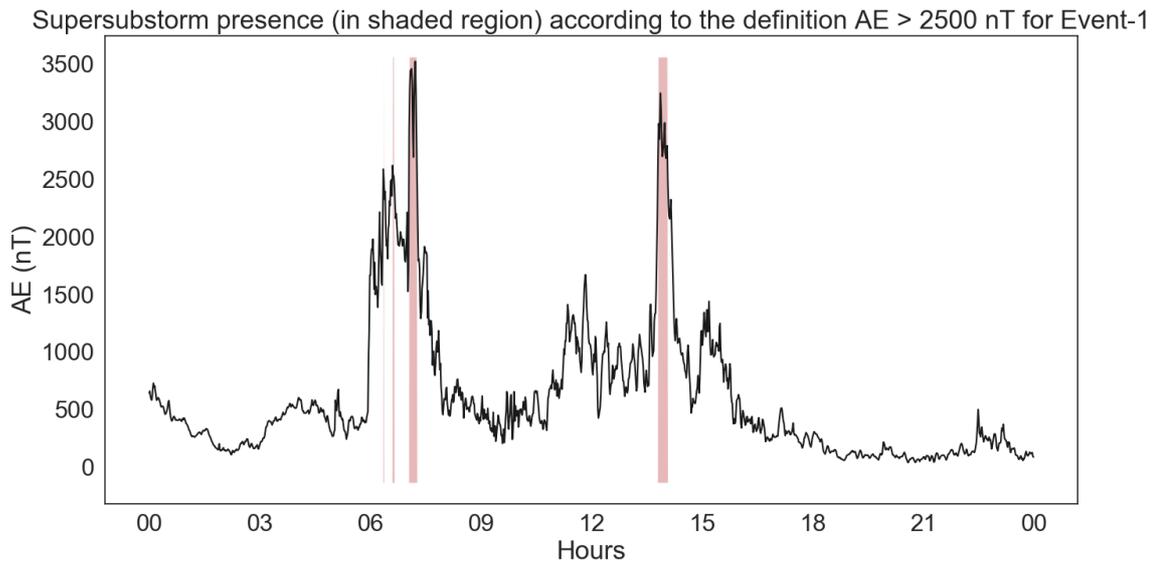

**Figure 4.2** Presence of Supersubstorm in the day of 24 November 2001. Supersubstorm is defined as the period when AE > 2500 nT.

**Event-2 Qualitative Analysis: Supersubstorm on 21 January 2005**

The IMF parameters, geo-magnetic indices, and the computed energy for this day is shown in Figure 4.3. The first panel shows the solar wind speed measured for the day. At around 17:12 UT, there is a sharp rise in the solar wind speed. In a span of about 12 minutes (from around 17:12 to 17:24), an increase in the solar wind speed from a level of 572 m/s to a level of 872 m/s is seen. There is a slow but continual rise in the solar wind speed which reaches a maximum level of 974.4 m/s at 19:43. Towards the end of the event, the solar wind speed is slightly reduced (though still higher than compared to that during the start of the day). Overall two phases in the wind speed can be seen, one low speed phase in the start with an average wind speed of 570.26 m/s and the high-speed phase towards the end of the day with an average wind speed of 912.07 m/s.

At around the same time that the increase in the solar wind speed was seen, change in other IMF parameters and geomagnetic indices are seen. The value of southward component of the magnetic field (Bz) is shown in the 2$^{nd}$ panel. Bz starts fluctuating from its stable range of 0 nT at around 17:12 UT. With a slight decrease up until 17:40 UT, it starts to rise with only local rise and fall until 18:16 UT. After this time, Bz drops sharply reaching its lowest value of -27.97 nT at 18:39 UT. Between 18:40 UT and 18:48 UT, the value of Bz recovers to a high value of 30.23 nT at 18:43 before falling to the value of around 17.8 nT again. After this, the value of Bz rises over the duration of the day, only showing minor variations along the way. The electric field component (Ey) plotted in the 3$^{rd}$ panel show a mirror-like activity as compared to Bz. The lowest value of Ey is -28.63 mV/m which occurs at 18:44 UT. The symmetric component of the ring current (SYM-H) is plotted in the 6$^{th}$ panel where the general behavior can be seen as the SYM-H value rising from its stable phased to reduce again slightly, rise again briefly to only continuously fall over the end of the day. The rise in SYM-H starts at around 17:10 UT where the SYM-H begins to rise to a local high value of 56.8 nT at 17:22 UT. The SYM-H value falls to a low of -41.86 nT at 18:40 to a peak of 45.8 nT at 18:57 UT. From this peak, the SYM-H fall gradually to an average level of ~ -70 nT towards the end of the day. The value of AE index is plotted in the 7$^{th}$ panel which shows on a broader sense a similar pattern as that of SYM-H. There is a rise in the AE-index value around the time when the increase in the solar wind speed was seen. This rise grows to a maximum of 3504 nT at 17:38 UT. From this high, the AE falls gradually towards the value of ~225 nT at the end of the day, with locals highs and lows along this fall.

The solar wind energy for this day is plotted in the 5$^{th}$ panel. The energy rises to the maximum of 0.68 * 10^14 joules/sec seen at in two steps. In the first step, the energy increases from an average level of 0.03 * 10^14 joules/sec to an average level of 0.25 * 10^14 joules/sec. This increase occurs at around 17:12 UT. The second step increase towards the maximum energy level starts at around 18:44 UT. From the maximum energy level, the energy value falls gradually towards the end of the day setting to an average value of ~0.25 joules/sec. The Akasofu parameter (epsilon) is plotted in the 4th panel. With a slight increase in the epsilon value seen starting at around 17:12 UT, there is a rapid increase phase which starts at 18:17 UT. The maximum of epsilon is 0.26 * 10^14 joules/sec which is reached at 18:39 UT. This sees a rapid fall corresponding exactly to the phase when there is a sharp increase in Bz, before rising again to the value of 0.18 * 10^14 joules/sec at 18:49 UT. From this level, the value of epsilon decreases to the average floor level of 0.002 * 10^14 joules/sec with only small duration increases seen e.g. at 20:38 UT and 21.06 UT.

The supersubstorm activity during this day according to the definition based on the AE index is shown in Figure 4.4. In the plot, the supersubstorm phase is shown by the shaded region. It can be ascertained that a single significant supersubstorm activity was seen. This activity started at 17:31 UT and ended at 17:53 UT. During this phase, the AE reached to the highest level of 3504 nT.

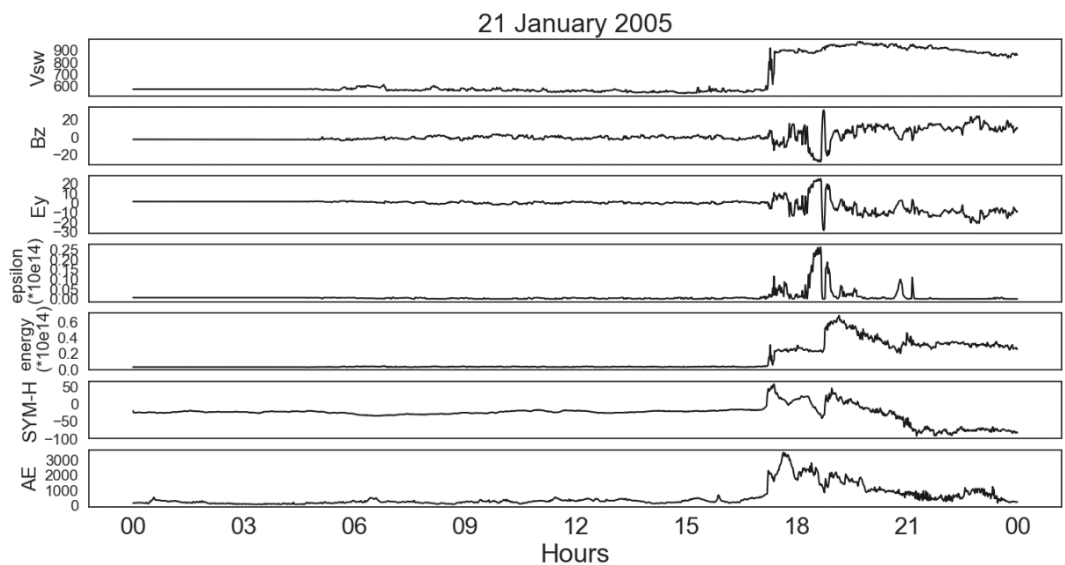

**Figure 4.3** IMF parameters and energy computed for Event-2 on 21 January 2005

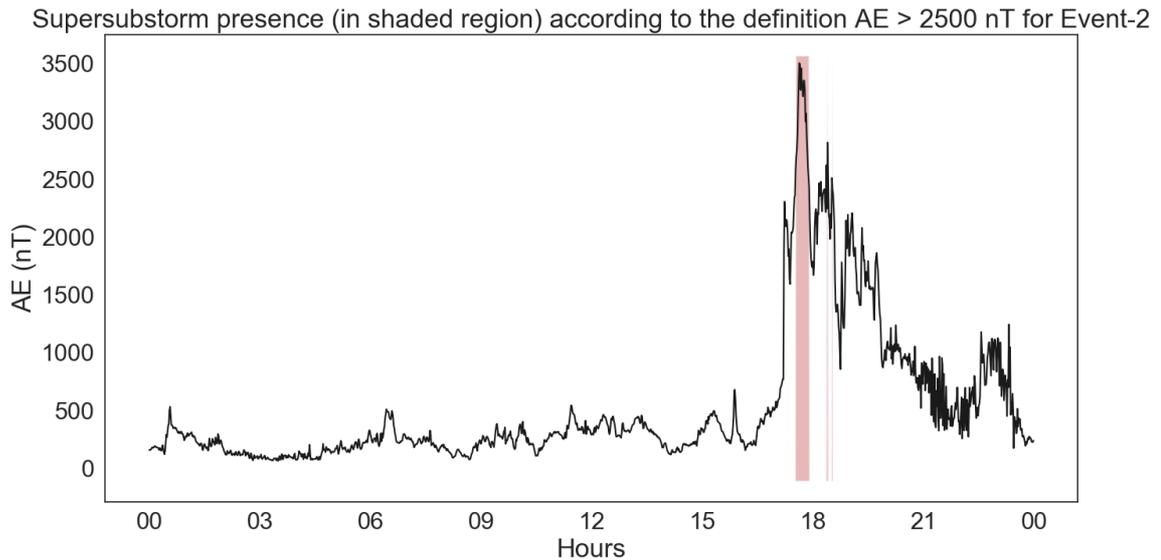

**Figure 4.4** Presence of Supersubstorm in the day of Event-2 on 21 January 2005. Supersubstorm is defined as the period when AE > 2500 nT.

**<u>Event-3 Qualitative Analysis: Supersubstorm on 24 August 2005</u>**

The IMF parameters, geo-magnetic indices, and the computed energy for this day is shown in Figure 4.5. The solar wind speed is shown in the first panel. The speed increases over the day in a step-like increases. There is a total of five steps counting the initial step with an average solar wind speed of around 410 km/s. This increases to the solar wind speed range of 450 km/s starting at 01:06 UT. This value continues until 06:18 UT when the solar wind speed then increases to ~530km/s. The fourth step occurs around 09:02 UT when the solar wind speed increases to 620 km/s. There is a missing data in the period between 11:57 UT and 16:13 UT. After the data is available, it can be seen that the solar wind speed is still maintaining the level of ~630 km/s. The final rise for the day in the solar wind speed is seen at 20:56 UT when the speed rises to the level of ~ 715 km/s. The southward component of the magnetic field (Bz) is shown in the plot in the 2$^{nd}$ panel. The value of Bz sees a large drop around 09:08 UT when it falls to the overall minimum of -55.57 nT at 10:02 UT. The value of Bz before this fall at 43.65 nT is the highest value of Bz recorded during this day. The nominal value of Bz for this day is in between the -2 nT to 2 nT range which is maintained both at the start of the day and towards the end. The electric field component (Ey) is plotted in the 3$^{rd}$ panel and shows again a similar but inverted pattern as Bz. The maximum and minimum of Ey is 34.1 mV/m and -24.87 mV/m reached at 10:03 UT and at 09:08 UT respectively. The SYM-H component is plotted in the 6th panel and the overall pattern can be described as one gradual fall from an initial level during the start of the day to the value maintained towards the end of the day. The highest value of SYM-H is 74 nT seen at 09:02 UT. From this point, SYM-H falls towards the lowest level of -179 nT seen at 11:50 UT. Then the SYM-H recovers slightly towards a stable level of ~ -100 nT for the rest of the day.

The solar wind energy for this event in shown in the 5th panel. The energy level rises from a level of around 0.04 * 10^14 joules/sec to the level of 0.1 * 10^14 joules/sec at 06:18 UT. Then a drop to a level of around 0.08 * 10^14 joules/sec is seen at 09:10 UT which roughly corresponds to the phase of decrease in Bz and second rise in solar wind speed. The energy rises again at 10:04 UT upto a maximum of 0.24 * 10^14 joules/sec. Towards the end of the day, there is a gradual decrease in the energy to finally settle around the level of 0.03 * 10^14 joules/sec. The Akasofu parameter epsilon is shown in the 4th panel. As a general description of the observed behavior of the epsilon, it can be said that a rise of epsilon can be seen at 09:07 UT where it rises from a level of 0.03 * 10^14 joules/sec. The rise reaches a maximum of 0.82 * 10^14 joules/sec at 10:04 UT. After that, the epsilon falls rather rapidly reaching to the similar levels as it were during the start of the day. This phase in the rise of the epsilon roughly coincides with the phase where a large dip in the value of Bz is seen.

The supersubstorm activity during this day according to the definition based on the AE index is shown in Figure 4.6. The supersubstorm phase is shown with the shaded region. Two closely located regions of subpersubstorm activity can be seen. The first one occurs at 10:07 UT and lasts for 13 minutes until 10:20 UT (barring tiny drops of AE just below the threshold in two of the 13 minutes of the measurements). The second closely located supersubstorm activity starts at 10:32 UT and ends at 10:40 UT, thus lasting for 8 minutes.

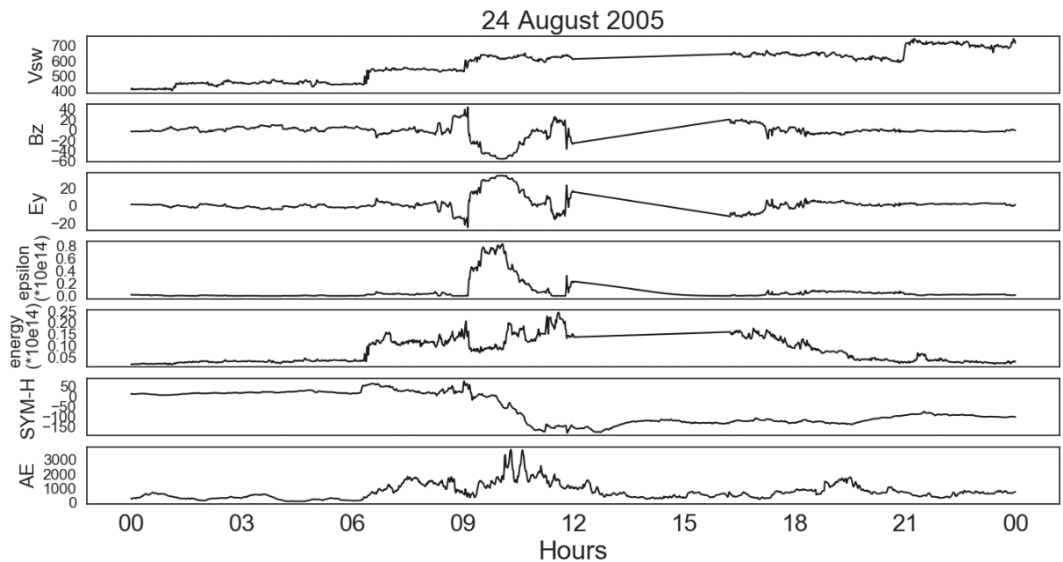

**Figure 4.5** IMF parameters and energy computed for Event-3 on 24 August 2005

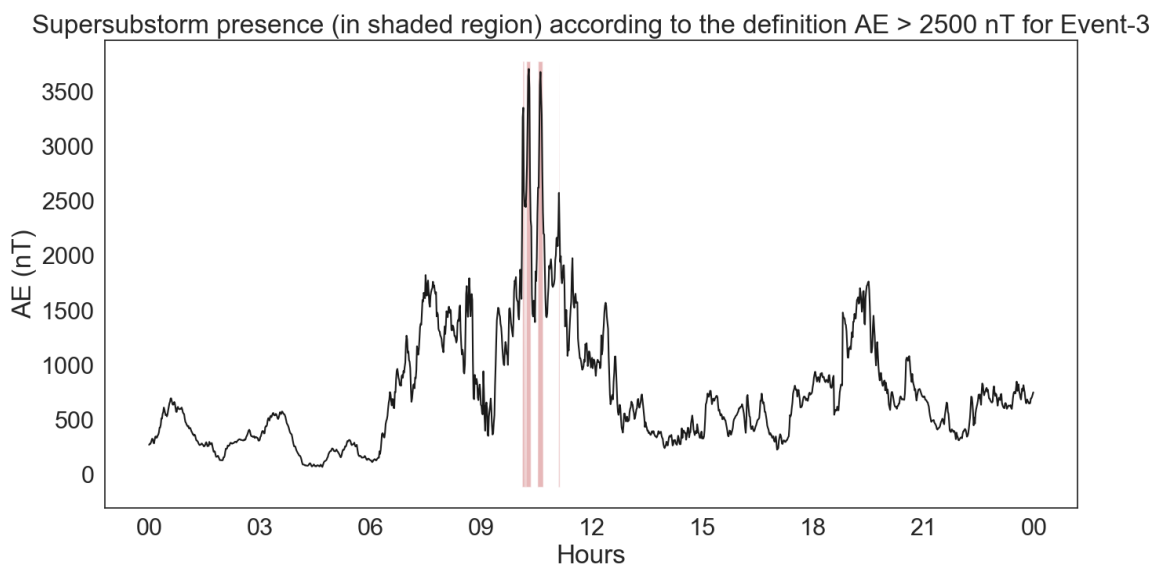

**Figure 4.6** Presence of Supersubstorm in the day of Event-3 on 24 August 2005. Supersubstorm is defined as the period when AE > 2500 nT.

**Event-4 Qualitative Analysis: Quiet day on 19 July 2006**

The IMF parameters, geo-magnetic indices, and the computed energy for this quiet day shown in Figure 4.7. It is to be noted that the Event-4 represents a quiet day. This is also reflected in the plots. The solar wind speed in plotted the first panel and its value varies only in a small range (285 km/s to 310 km/s). On this day, the solar wind speed gradually decreases from the average level of 300 km/s towards the level of 285 km/s during the mid-day before rising gradually back to the level of 300 km/s at the end of the day. For both the Bz (2nd panel) and the Ey (3rd panel) value, there is very less activity during the day starting from 06:10 UT up until 17:35 UT. Before and after this time, there are small deviations observed but only by a small margin. For example, Bz drops from 0 nT during the start of the day to below -2 nT before rising back to the level of 2 nT. Similar rises and falls in the value of Bz, within +/- 2 nT is also seen towards the end of the day. Variations in Ey (3rd panel) are constrained within +1 mV/m to -0.5 mV/m. The symmetric component of the ring current (SYM-H) is shown in the 6th panel. The values of SYM-H rises from the level of -2.5 nT to around 2.5 nT at around 06:50 UT before falling back to the level of -2.9 nT 16:10 UT. It rises again towards 9 nT at 20:45 UT and falls slightly towards 0 nT – 2.5 nT range at the end of the day. The AE index (7th panel) only shows a rise and fall activity for a short time between 02:00 UT and 05:00 UT. Otherwise its value is generally in the level of 45 nT – 50 nT for the major part of the day. Overall variation is also quite low with the maximum value of 182 nT reached at 03:37 UT.

The solar wind energy (plotted in the 5th panel) also only minor deviations. In the start of the day, it falls from the level of 0.016 * 10^14 joules/sec toward the lowest level for the day of 0.009 * 10^14 joules/sec at 09:00 UT. It rises again at around 17:30 UT towards the level of 0.018 * 10^14 joules/sec. The epsilon parameter (shown in the 4th panel) shows some variability from the start of the day up until 06:17 UT where it settles towards a level of 0.00004 * 10^14 joules/sec. During this initial variability phase, the maximum of 0.0025 * 10^14 joules/sec at 02:49 UT.

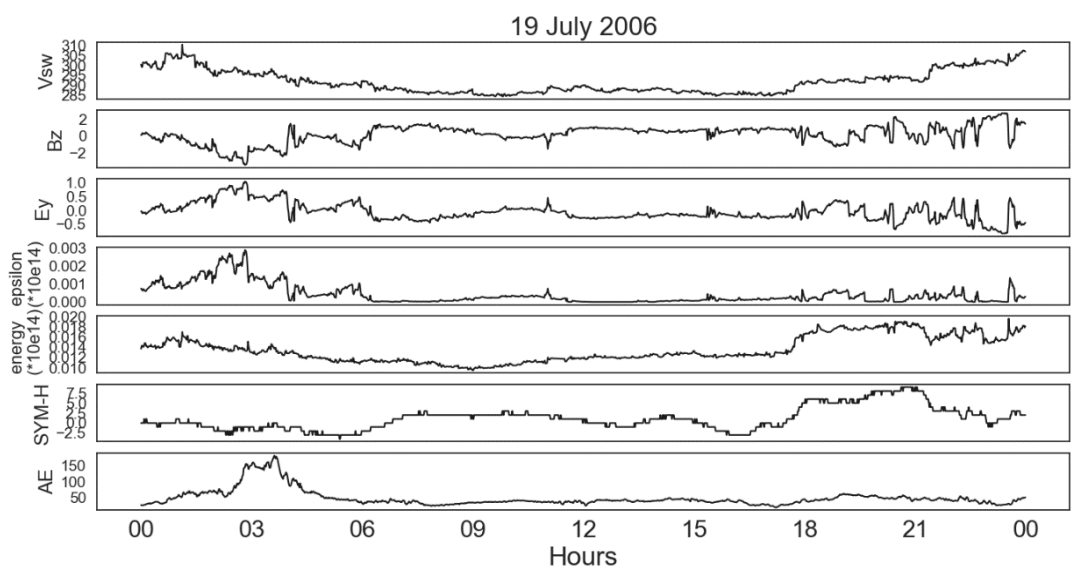

**Figure 4.7** IMF parameters and energy computed during quiet day on 19 July 2006

## 4.2 Correlation of energy parameters with the geo-magnetic indices

In the previous section, a qualitative analysis of the observed pattern of solar wind energy, epsilon parameter, IMF parameters and different geo-magnetic indices for different event days were presented. Here we present the correlation between the energy parameters and other IMF parameters.

Correlation coefficient between the energy parameters (solar wind energy and epsilon parameter) with the geo-magnetic indices were computed. The summary of these correlations are shown in Section 4.3.

**Table 4.1** Correlation of solar wind energy and Akasofu parameter (epsilon) with other geo-magnetic indices. Pearson correlation coefficient is used to obtain the correlation from the data points from all four events considered in this work.

| Correlation With | Solar Wind Energy | Epsilon |
|---|---|---|
| **Bz** | 0.34 | -0.95 |
| **SYM-H** | -0.34 | -0.01 |
| **AE** | -0.02 | 0.09 |

Scatter plot showing the relation between energy parameters and the geomagnetic indices/IMF parameter are also shown. The scatter plot between solar wind energy and IMF-Bz is shown in Figure 4.8. A slight positive correlation is obtained. A higher correlation pattern can be seen specially for the higher energy region (right of the plot), though only few data points are present in this region. The relation between solar wind energy and SYM-H can be seen from Figure 4.9 where it can be seen that there is a slight negative correlation between these. At higher energy levels (right of the plot), almost a flat line relation between the energy and SYM-H can be seen which indicates no correlation between these for high energy levels. A low correlation (~0) between solar wind energy and AE index, as seen in Figure 4.10, shows that the AE index is independent of the energy levels.

Epsilon and IMF-Bz has a very high negative correlation. With increase in epsilon, a corresponding decrease in IMF-Bz can be seen and vice versa. This strong relationship can be explained by observing the formula to compute epsilon. IMF-Bz appears through the angle parameter theta when computing epsilon. However no relationship between epsilon and SYM-H (Figure 4.12) as well as epsilon and AE-index (Figure 4.13) can be seen, as evident from a very low correlation close to zero.

One of the most interesting insights from the correlation analysis can be seen from Figure 4.14. The figure shows the relation between solar wind energy and the Akasofu parameter epsilon. A moderately strong correlation, albeit in the negative direction, of -0.40 is seen. When the solar wind energy value is low, the obtained epsilon values are one of the highest. This means that the coupling of the solar wind energy to the magnetosphere is highest when the solar wind energy level themselves are lower. With increase in the solar wind energy level, the value of epsilon reduces in general indicating lowered coupling of the energy to magnetosphere. It seems as if the epsilon saturates to particular levels, around $10^{12}$ joules/sec, when the solar wind energy is very high.

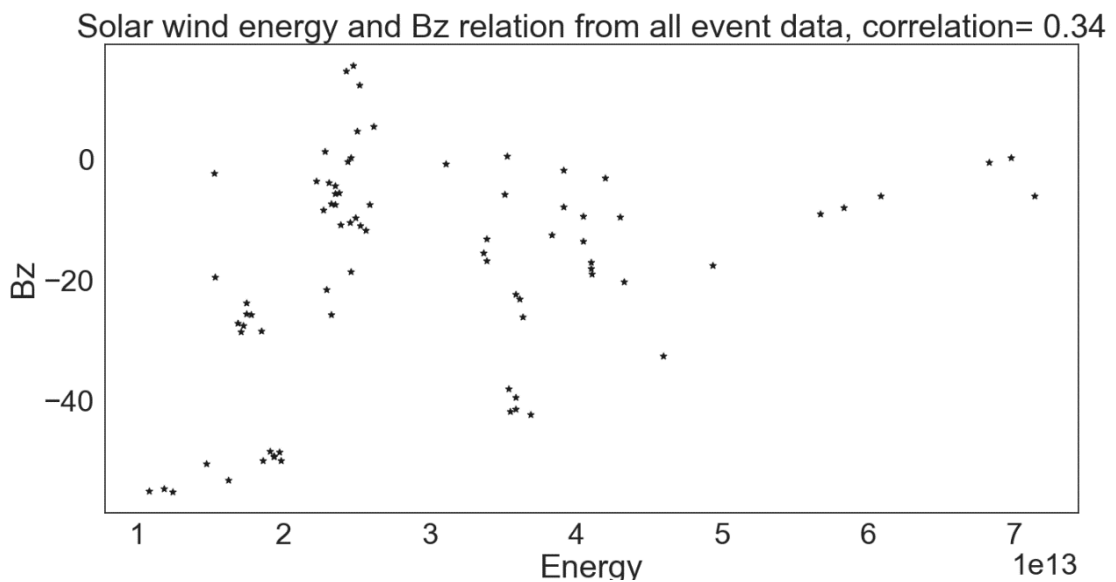

**Figure 4.8** Relation between solar wind energy and IMF-Bz (southward component of IMF) with data from all the event days (24 November, 2001; 21 January, 2005; 24 August, 2005; 19 July, 2006. )

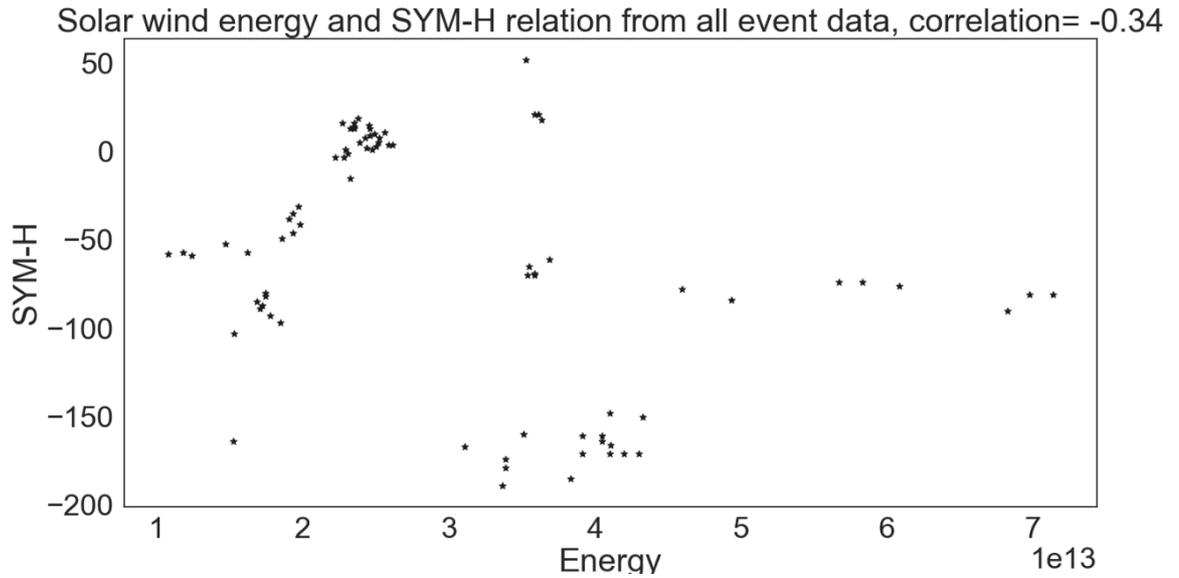

**Figure 4.9** Relation between solar wind energy and SYM-H index (symmetric component of the ring current) with data from all the event days.

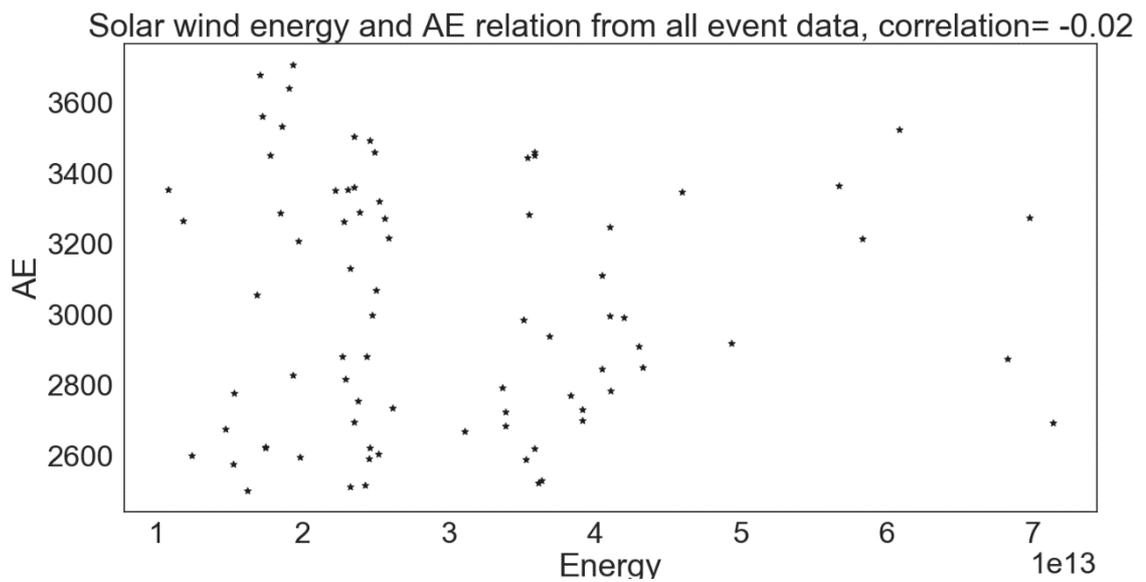

**Figure 4.10** Relation between solar wind energy and the AE index from the data obtained from all event days (24 November, 2001; 21 January, 2005; 24 August, 2005; 19 July, 2006.)

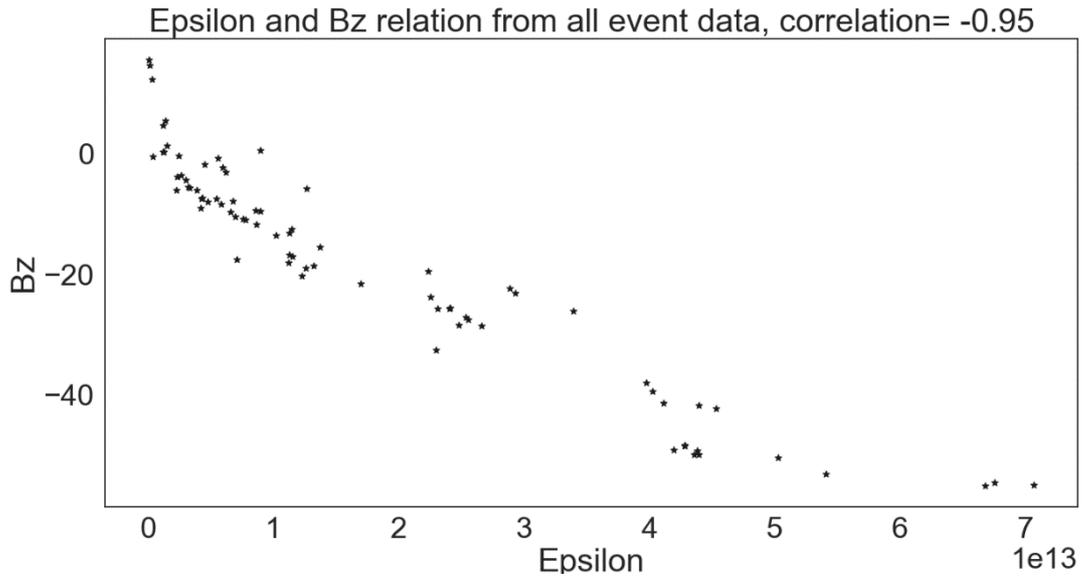

**Figure 4.11** Relation between epsilon and Bz (southward component of IMF) from data of all event days (24 November, 2001; 21 January, 2005; 24 August, 2005; 19 July, 2006.)

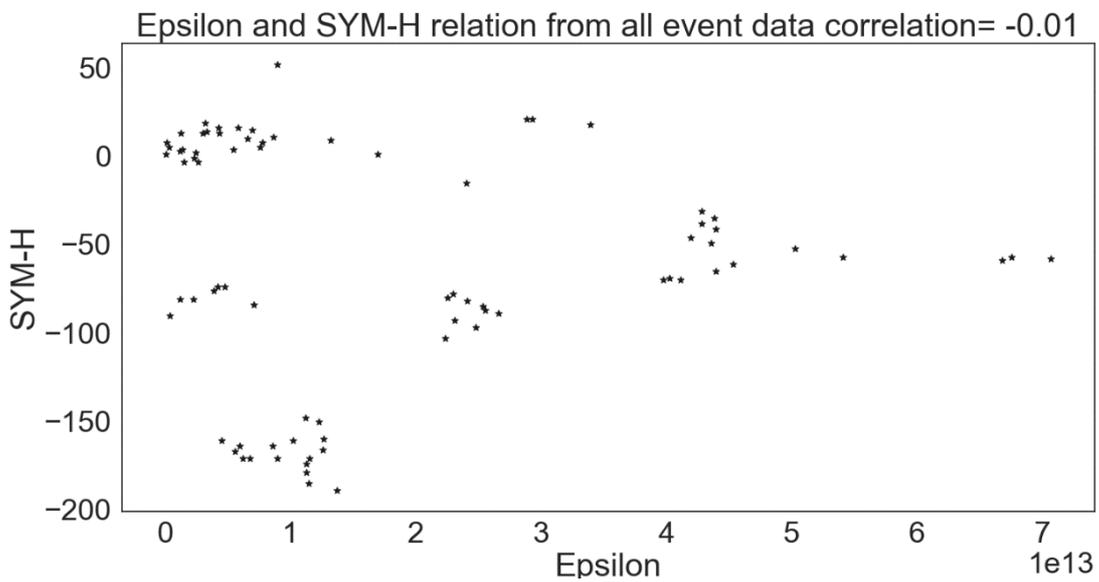

**Figure 4.12** Relation between epsilon and SYM-H (symmetric component of ring current) from data of all event days (24 November, 2001; 21 January, 2005; 24 August, 2005; 19 July, 2006.)

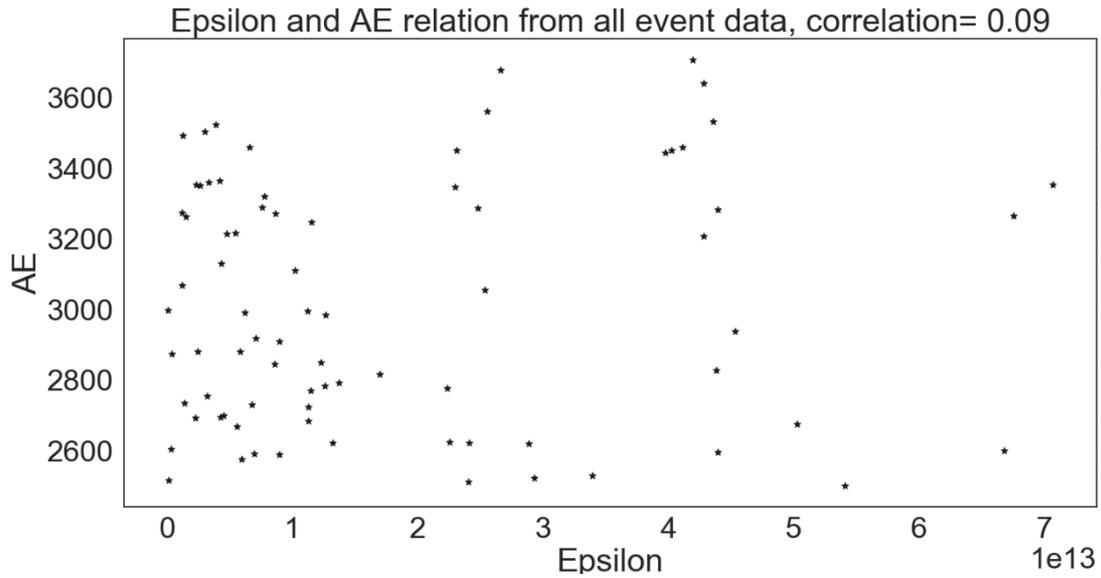

**Figure 4.13** Relation between epsilon and AE-index as obtained from observations from all considered events (24 November, 2001; 21 January, 2005; 24 August, 2005; 19 July, 2006.)

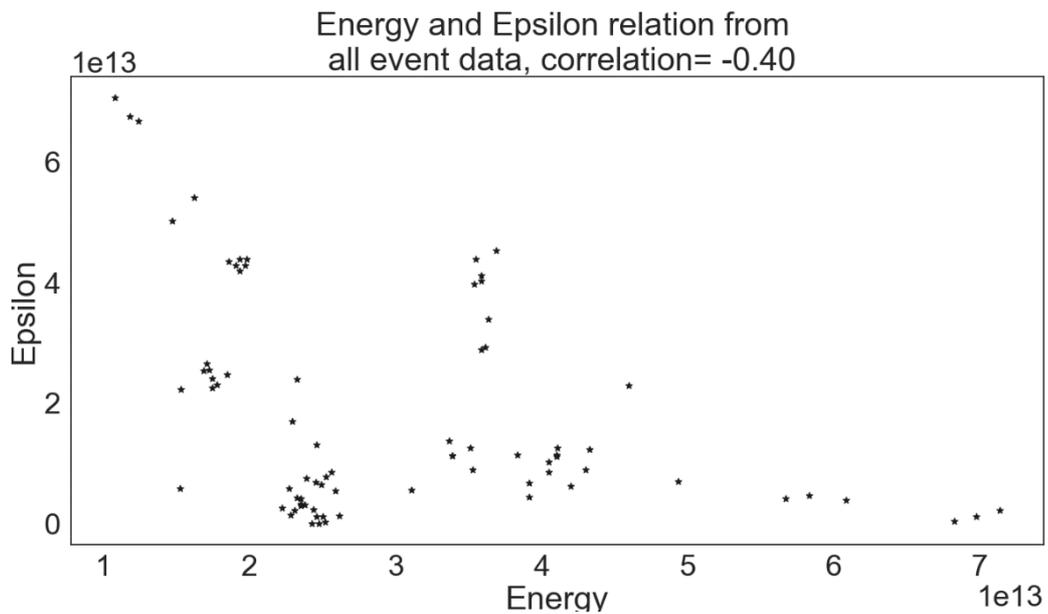

**Figure 4.14** Relation between solar wind energy and epsilon observed from data points of all of the considered four events (24 November, 2001; 21 January, 2005; 24 August, 2005; 19 July, 2006.)

## 4.3   Cross-correlation between energy and the geo-magnetic indices

In the previous section, we reported the correlation between energy parameters and the geo-magnetic indices. This analysis helped uncover the dependency and relation between energy parameters and other indices. However, a correlation analysis with a correlation coefficient cannot fully capture the phase delay relation between different signal. For example, an effect of increase energy level might appear as change in one of the geo-magnetic indices only after a certain time. This effect might not be fully captured by the analysis using correlation coefficient alone. In this section, we study and report the cross-correlation of energy (solar wind energy and epsilon parameter) and other geo-magnetic indices. This analysis is done individually for each event.

The cross-correlation analysis of solar wind energy and IMF-Bz, SYM-H, and AE index for 24 November 2001 (Event-1) is shown in Figure 4.15. The relation of solar wind energy and AE index is overall a low positive, with maximum correlation reaching slightly higher than 0.25. The correlation between energy and SYM-H is strongest (albeit in the negative direction) at -0.87 for a lag corresponding to -362 minutes. For this particular event, there is a rise in solar wind energy which leads to a fall in SYM-H with a bit of delay which leads to this correlation at a delay. The correlation between the energy and the AE-index is highest at 0.8 with no lag. Therefore, the impact of solar wind energy is seen immediately in the AE-index for this event day.

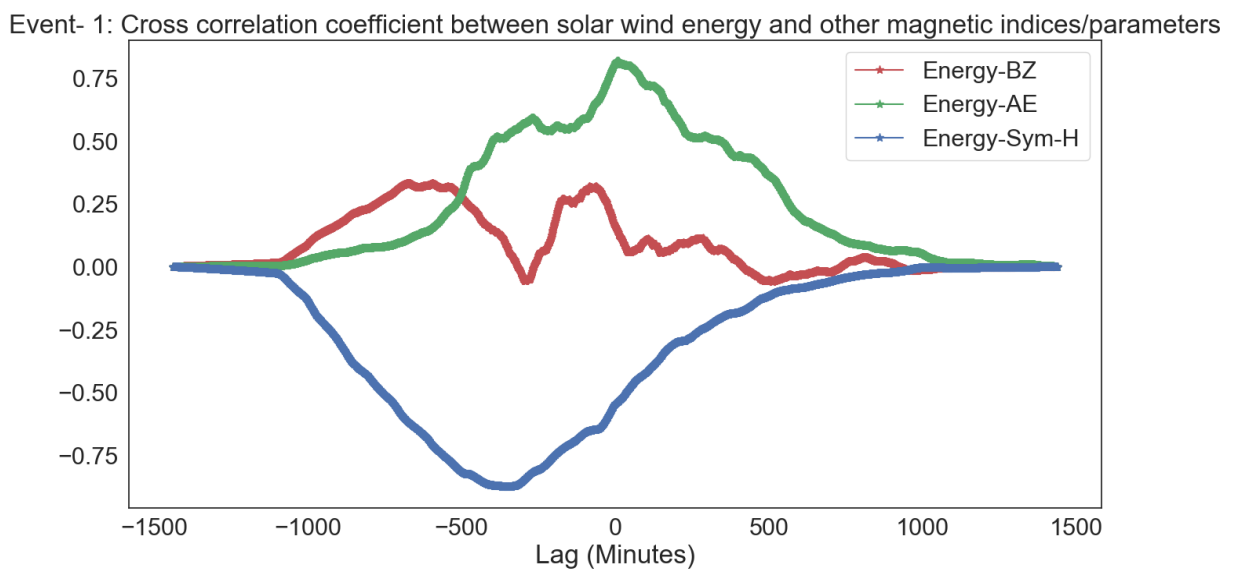

**Figure 4.15** Cross correlation between solar wind energy and other geomagnetic indices for 24 November 2001 (Event-1)

Cross-correlation analysis for 21 January 2005 (Event-2) is shown in Figure 4.16. For this event, a bit higher positive correlation between energy and IMF-Bz is seen with a maximum of 0.55 at a lag of -85 minutes. The SYM-H index correlates negatively, with a highest correlation of -0.7 at the lag of -140 minutes. The AE-index is correlated with the energy with almost no lag. The highest correlation reached is close to 0.89. This also shows again for this event that the energy shows a direct impact on the AE index while a lagged relationship exists between energy and the SYM-H parameter.

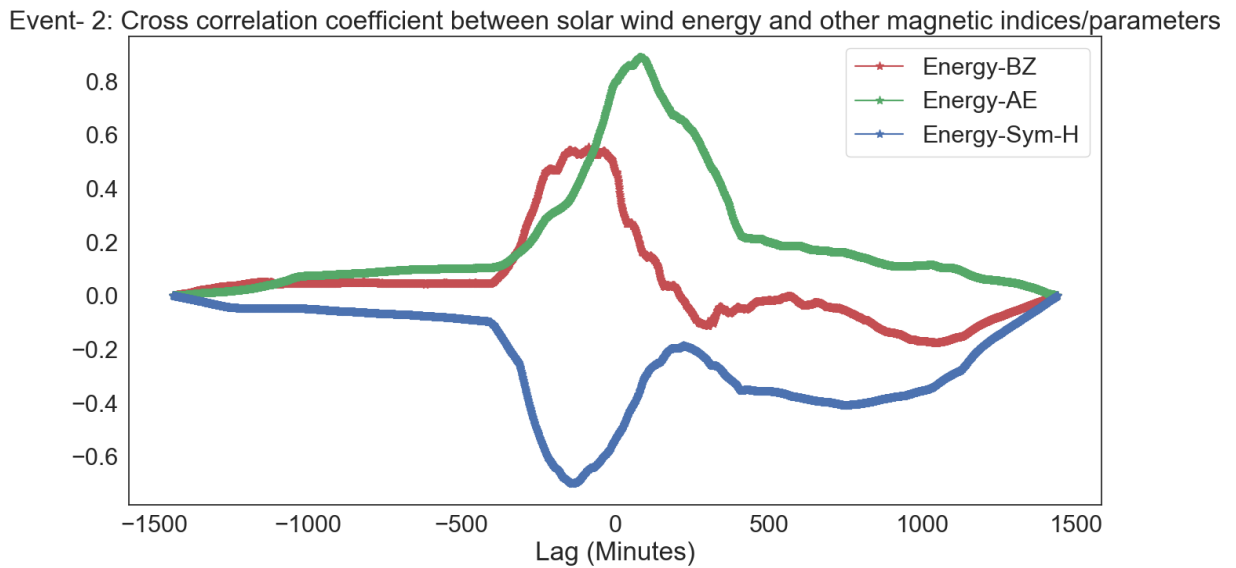

**Figure 4.16** Cross correlation between solar wind energy and other geomagnetic indices for 21 January 2005 (Event-2)

Cross-correlation analysis for 24 August, 2005 (Event-3) is shown in Figure 4.17. No significant correlation is obtained between solar wind energy and Bz, even at positive and negative lags. The relation of energy and SYM-H reaches the strongest correlation (negative side) of -0.9 at -240-minute lag. As for the AE-index, the relation is strongest with a positive correlation in the range of 0.8 at no lags.

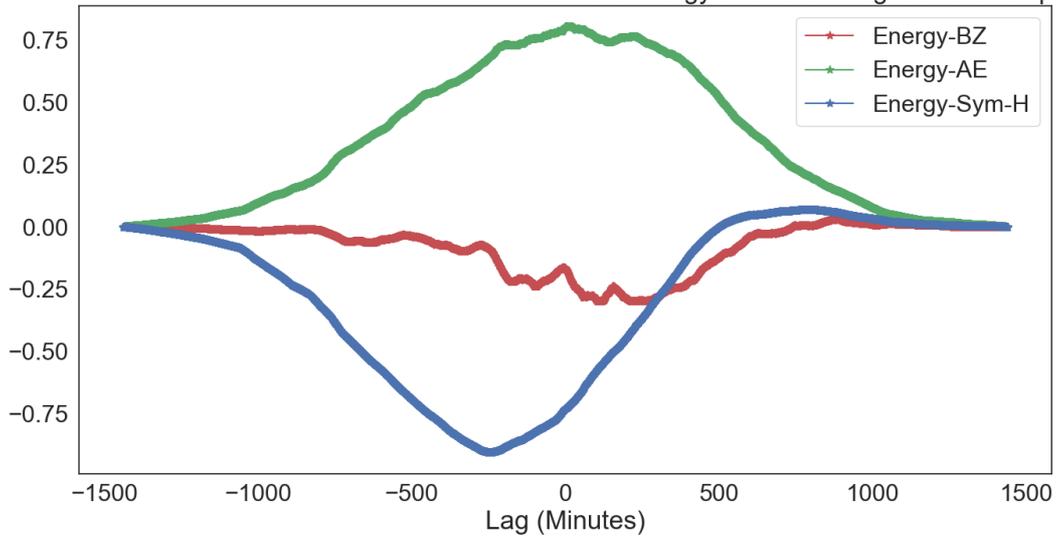

**Figure 4.17** Cross correlation between solar wind energy and other geomagnetic indices for 24 August, 2005

Cross-correlation analysis for Event-4 is shown in Figure 4.18. It is to be noted that this event corresponds to a quiet day with no supersubstorm occurring on this day. The energy only correlates slightly with Bz, with a highest positive correlation reached to a range of 0.4. The energy also correlates similarly, at a lower level, with SYM-H index. This is in contrast to the observed behavior of the relation between energy and SYM-H for other events where supersubstorm occurred. Similarly, to the other events however, the energy correlates strongly and positively with the AE index. The correlation value reaches greater than 0.8 for no lag.

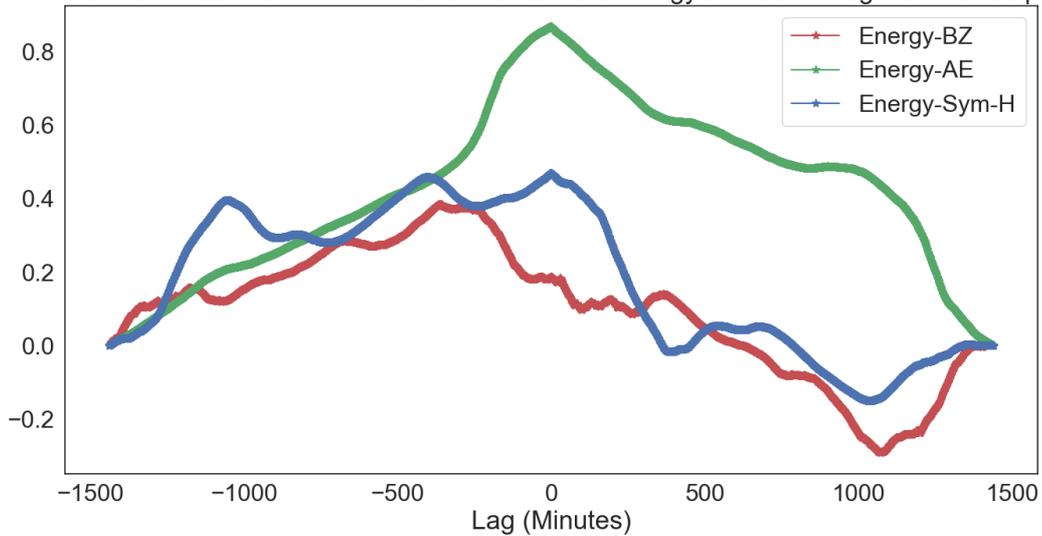

**Figure 4.18** Cross correlation between solar wind energy and other geomagnetic indices for 19 July 2006

The cross-correlation analysis of the epsilon parameter with other geomagnetic indices are shown in Figure 4.19, Figure 4.20, Figure 4.21, and Figure 4.22 for Event-1, Event-2, Event-3 and Event-4 respectively. The correlation of Bz and epsilon is low for all events with strongest correlation reaching only as high as 0.4. This behavior is similar to the correlation between solar wind energy and Bz. For events with supersubstorm, the epsilon correlates mildly strong in the negative direction with the SYM-H. The correlation is slightly weaker than that obtained between energy and SYM-H, however the correlation is still at a negative lag which is similar to what had been seen between energy and SYM-H. For 24 November, 2001 (Event-1), the strongest correlation of -0.54 is reached at the lag of –374 minutes. For Event-2, the strongest correlation of -0.51 is reached at the lag of –222 minutes. And for Event-3, the strongest correlation of -0.6 is reached at the lag of –157 minutes. AE-index correlates positively for no lag or for only a slight lag (~30 minutes). This shows that the epsilon parameter correlates almost directly (i.e. with little or no lag). The highest correlation values are 0.65, 0.62, 0.70, and 0.87 respectively for Event-1, Event-2, Event-3, and Event-4.

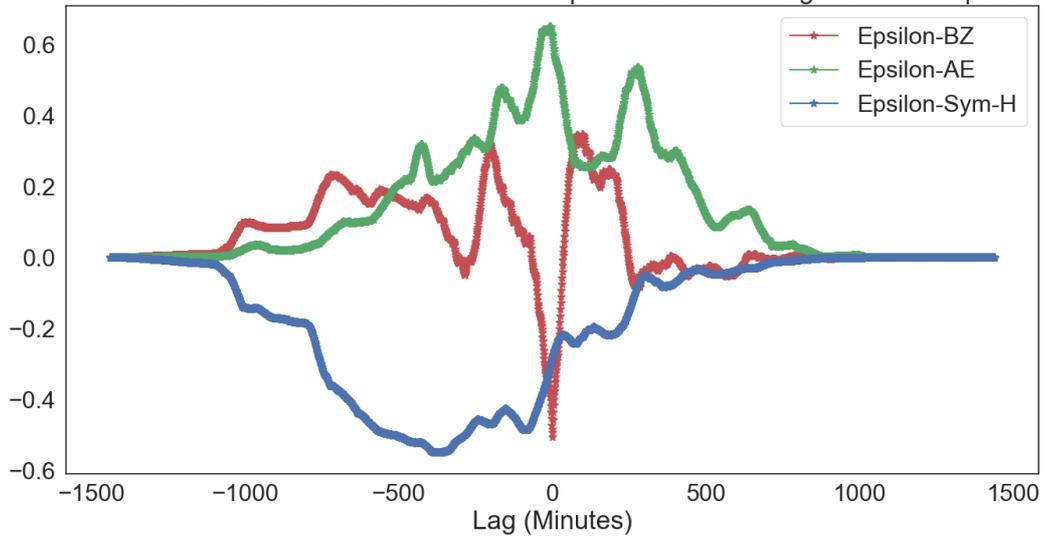

Figure 4.19 Cross correlation between epsilon and other geomagnetic indices for 24 November, 2001(Event-1)

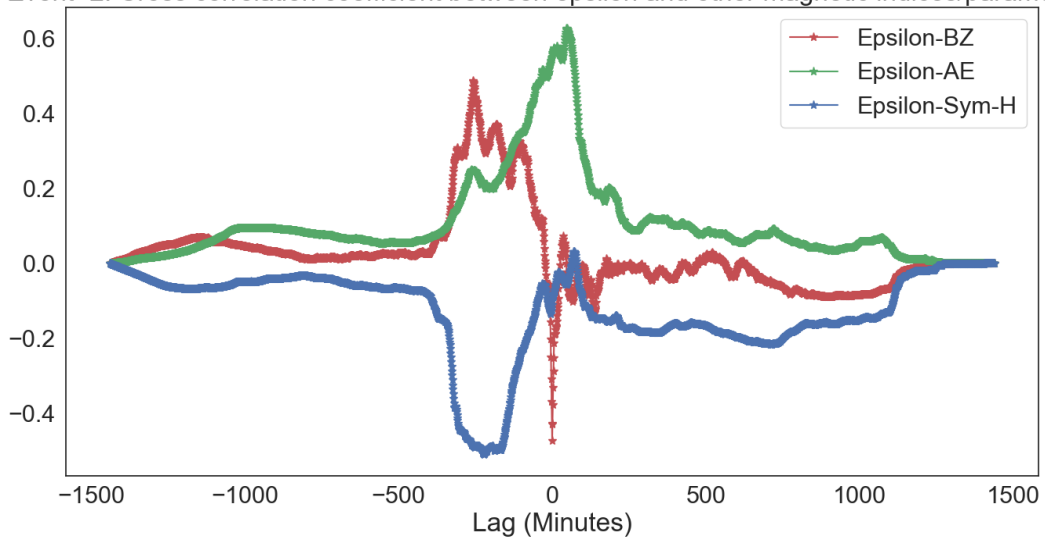

Figure 4.20 Cross correlation between epsilon and other geomagnetic indices for 21 January, 2005(Event-2)

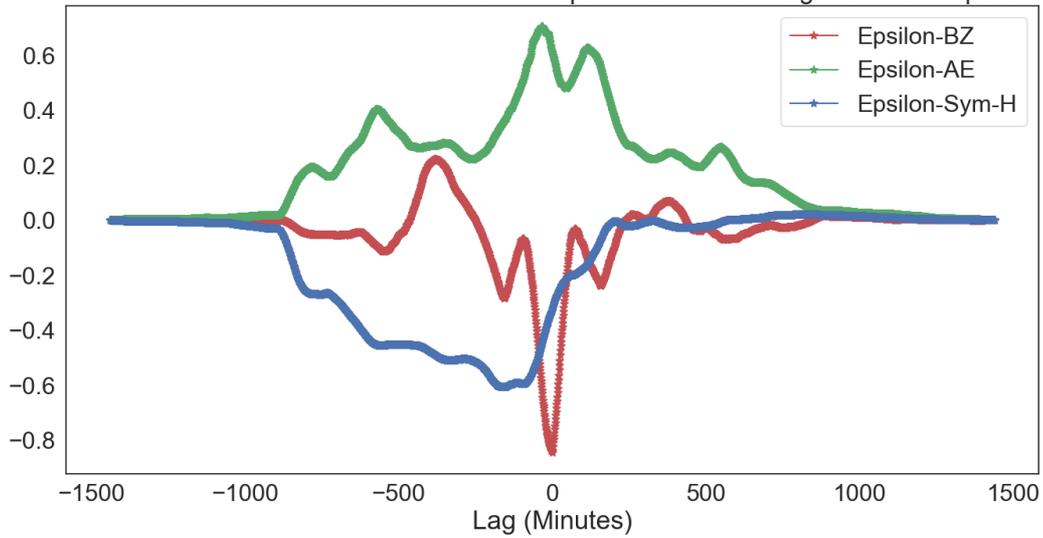

**Figure 4.21** Cross correlation between epsilon and other geomagnetic indices for 24 August, 2005 (Event-3)

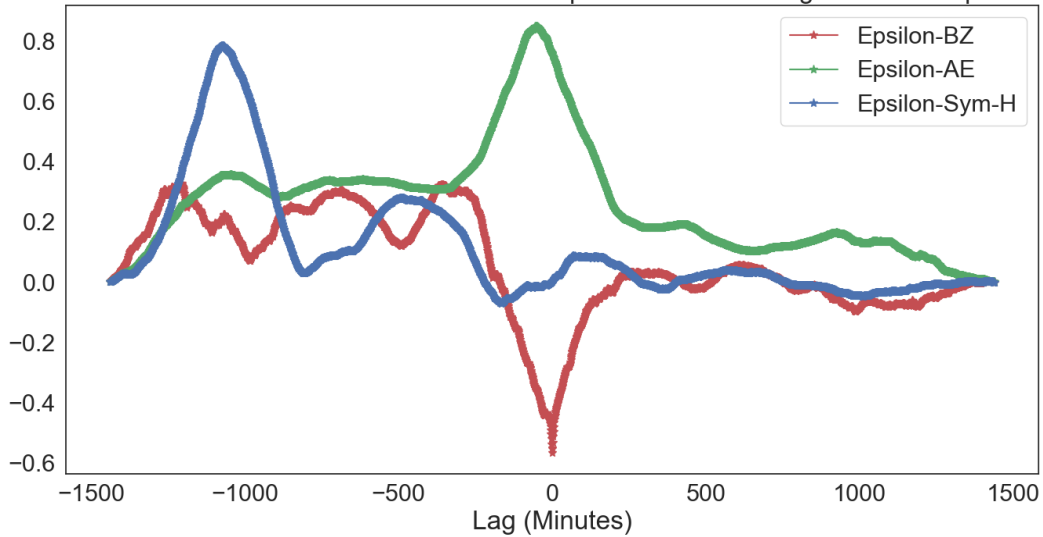

**Figure 4.22** Cross correlation between epsilon and other geomagnetic indices for 19 July 2006 Event-4

## 4.4 Summary of energy, epsilon parameter and their ratio during different events

In the previous sections, the qualitative analysis of energy parameters and geo-magnetic indices along with the correlation analysis to study the relationship among these were presented. Such analysis helped to understand the dynamics of different phenomenon during the supersubstorm event and put it in comparison to the dynamics of the quiet day. The correlation (including cross-correlation) further revealed the strength of relation (and any phase delay) between energy parameters and geo-magnetic indices.

In this section, we present the summary analysis (on an event level and overall) of the obtained results regarding the supersubstorm event (e.g. their duration), and the energy parameters computed. While the previous qualitative analysis and the correlation analysis helps to understand the dynamics, the summary analysis presented here helps to understand the global scenario during the considered event day (e.g. how long was the supersubstorm event, how much of the solar wind energy translates to the magnetospheric coupling as indicated by the Akasofu parameter, etc.). For example, understanding the total duration of supersubstorm event for the given event day helps to contextualize how much of the energy dynamics in a given day can be attributed to supersubstorm activity.

The duration of the supersubstorm event in the considered event days is shown in Figure 4.23. This duration has been computed from the duration when the AE index exceeded the threshold when it is considered to be supersubstorm. Therefore, the duration in itself is sensitive to the value of the threshold on AE chosen. The results that are shown here corresponds to our chosen threshold of 2500 nT. It can be seen that the highest duration of supersubstorm activity occurred for the 24 November, 2001 (Event-1). Here the duration was greater than 30 minutes in total. In comparison, the event duration for the supersubstorm event was around 25 minutes for Event-2 day and ~20 minutes for 24 August, 2005 (Event-3 day). Thus for a 24 hour period considered for analysis, the total duration of supersubstorm activity is only between 20-30 minutes, which as we know from our qualitative analysis are also not continuously occurring. E.g. there may be multiple closely occurring (possibly driven by same solar event) or far apart in time supersubstorm activities (possibly driven by different solar events).

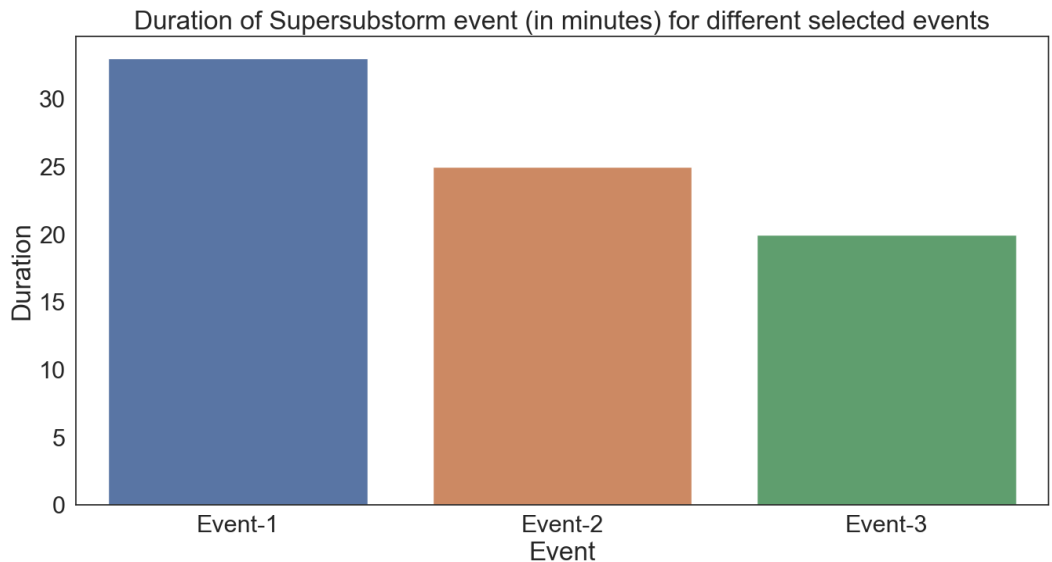

**Figure 4.23** The duration of supersubstorm as indicated by the AE index for the three different event days.

The average solar wind energy and the epsilon for each of the considered event days are shown in Figure 4.24. As the energy levels are low on Event-4 day in comparison to other event days, a separate zoomed plot is shown in Figure 4.25. The ratio between the average epsilon parameter and the average energy level, referred to as the energy ratio, is shown in Figure 4.26. The average energy is higher for Event-1 and succesively lower for Event-2, Event-3, and Event-4. The energy levels are the lowest for the quiet day (Event-4). Understandably, the average energy is higher for days with supersubstorm as it can be seen from this result too. An interesting insight is obtained when comparing the energy and epsilon values for the three events with supersubstorm. For the first and second event, while the average solar wind energy is higher ($0.16 * 10^{14}$ joules/sec and $0.12 * 10^{14}$ joules/sec respectively) the average epsilon values are lower in comparison ( around $0.04 * 10^{14}$ joules/sec and $0.01 * 10^{14}$ joules/sec respectively). In contrast to these, for Event-3 the average epsilon value is very close to the value of average solar wind energy. This might be indicative of the negative correlation observed between solar wind energy and epsilon (Figure 4.14). The dynamics between solar wind energy and epsilon is significant. It shows how much of the solar wind energy is coupled into the magnetosphere. This dynamics can also be shown by the ratio between epsilon and solar wind energy, which we refer to as energy ratio. Energy ratio for different event days are shown in Figure 4.26. As indicated earlier, the energy ratio is significantly higher on Event-3 day compared to other days with or without supersubstorm.

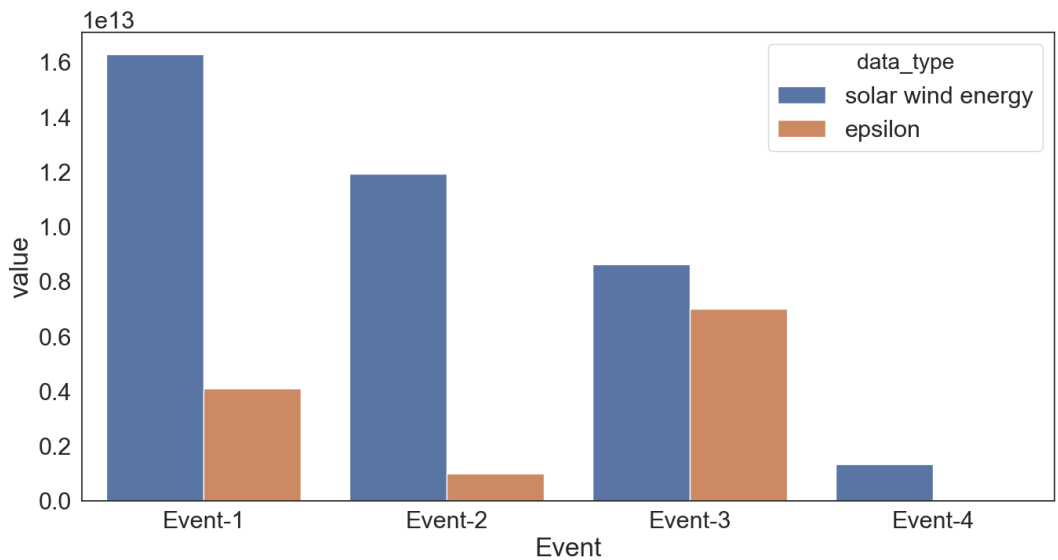

**Figure 4.24** Average solar wind energy and the Akasofu parameter (epsilon) for the event days considered.

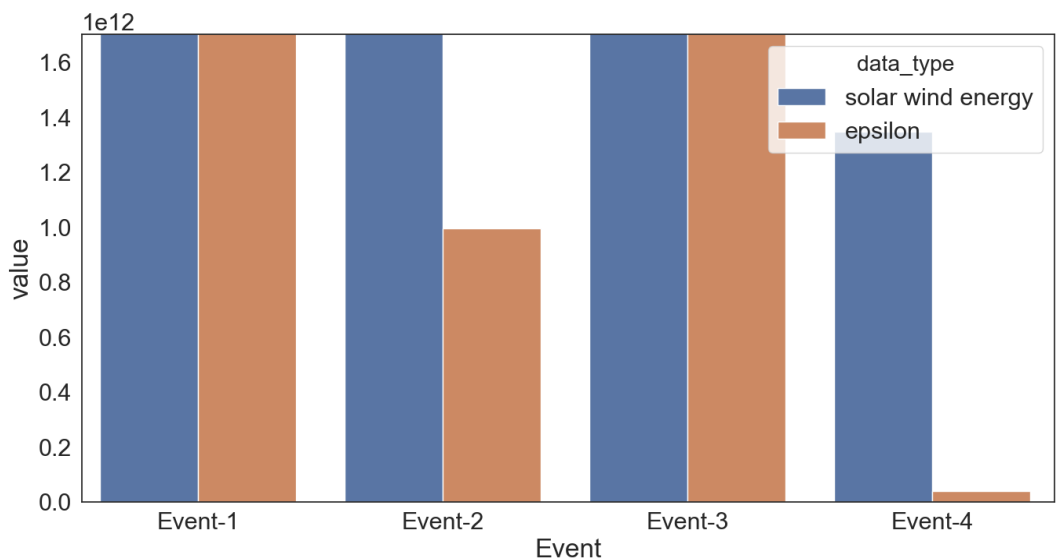

**Figure 4.25** Zoomed plot of Figure 4.24 showing Average solar wind energy and the Akasofu parameter (epsilon) for the event days considered.

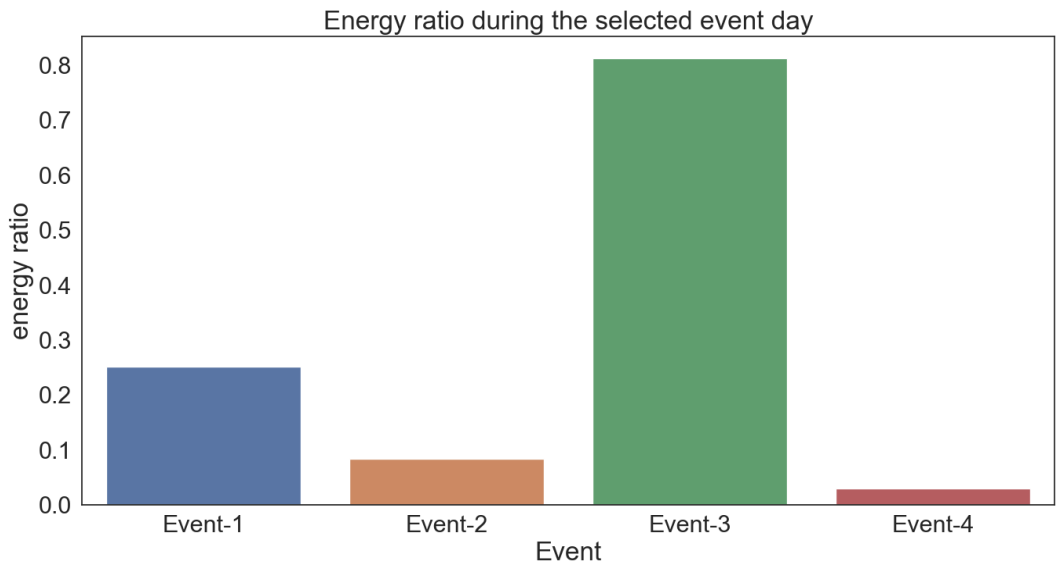

**Figure 4.26** Energy ratio, defined as the ratio of the average epsilon and the average solar wind energy for the four different event days.

The energy dynamics during the supersubstorm activity only is shown separately in Figure 4.27 with the average solar wind energy and average epsilon plots for each event day. Comparing the results in Figure 4.27 with that in Figure 4.24, it can be seen that the energy levels are quite high during the supersubstorm phase. E.g. for Event-1, during supersubstorm phase the energy levels are greater than 0.4 * 10^14 joules/sec while in the day itself the average energy level was 0.16 * 10^14 joules/sec. Thus the average energy during supersubstorm are four times that the energy level maintained during the day, as an example on this event day. The energy ratio for supersubstorm is in general higher when compared to the energy ratio seen for the observations from the whole day (compare the results in Figure 4.26 and Figure 4.28). One peculiarity seen from the results in Figure 4.27 and Figure 4.28 is the higher level of epsilon compared to the total solar wind energy for the supersubstorm phase of Event-3. This is not a plausible result as the solar wind energy should be larger than the coupled energy into the magnetosphere. However the supersubstorm represents observation from only few data points (in this case only few minutes) and this aberrations might be due to faulty observations/data points. The number of observations/data points available to compute the energy levels and/or epsilon parameter therefore seems to be important to be reported when reporting the values of the computed energy levels.

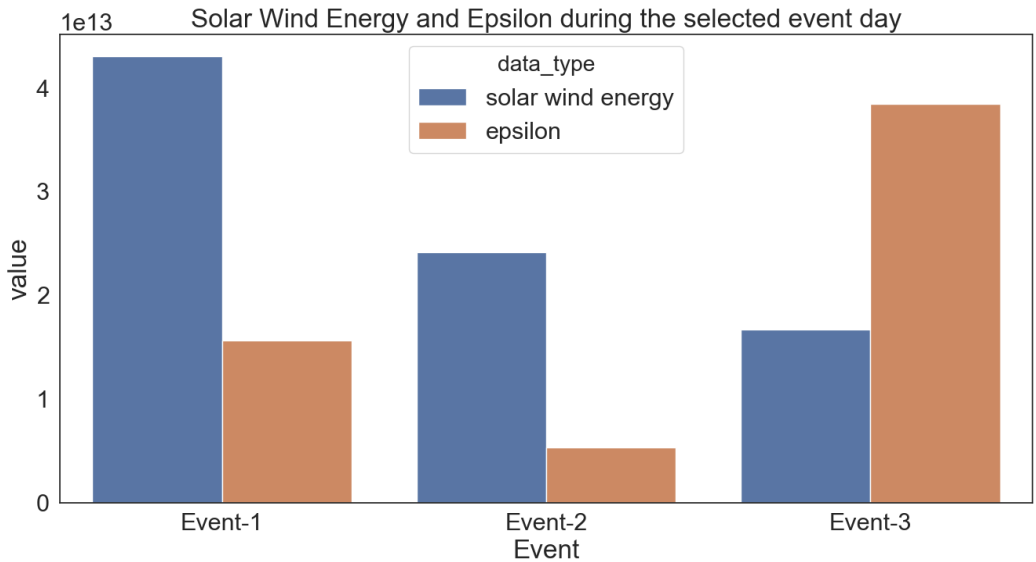

**Figure 4.27** Average Solar wind energy and epsilon parameter computed during the supersubstorm event.

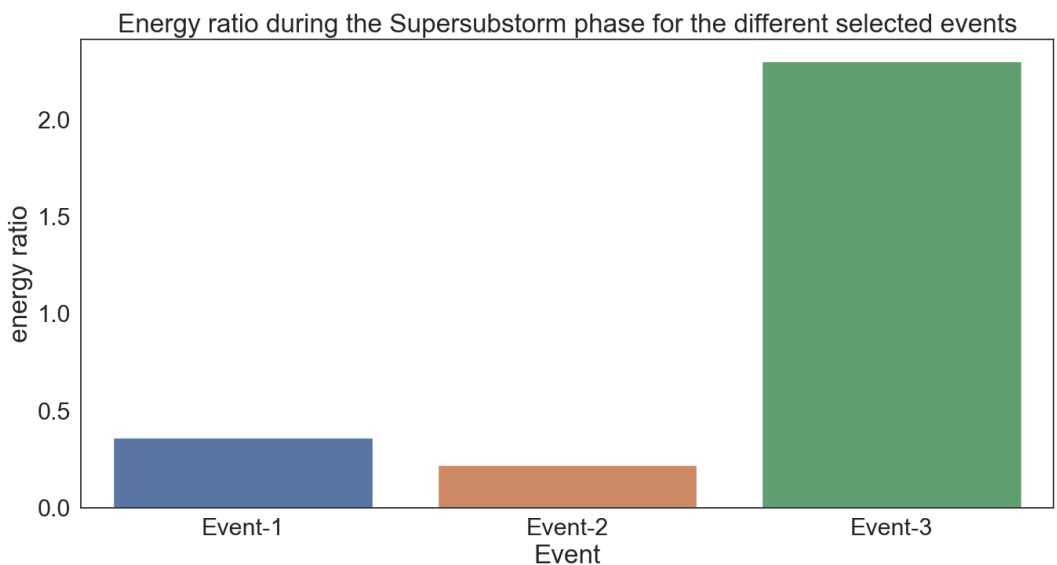

**Figure 4.28** Energy ratio, the ratio of the average solar wind energy and the average epsilon computed during the supersubstorm event.

Overall, from the three event days, we found that the supersubstorm phase in longest for Event-1 but this phase also lasts only about 30 minutes. Energy levels are higher in the days with supersubstorm compared to the quiet days. Also when computed separately, the energy contents in the supersubstorm phase is much higher than the average energy level.

# Chapter 5 Conclusion

Solar physics has significant impact on earth. Among different physical phenomenon arising from the activities of sun, supersubstorm is one of them which bring significant changes in the earth, for example by increasing the total magnetic field in the earth's surface. This effect in result might lead to consequences like disturbances in communication system, electric grid damage, hampering of the space communications etc. As the solar activity has significant impact on our solar system and in particular to earth, it is important to study and analyze solar activity induced events like geomagnetic storms and substorms.

In this thesis we explored the energy dynamics of supersubstorm by conducting a qualitative and quantitative analysis of computed energy for three days with supersubstorm and a control day without supersubstorm (quiet day). Solar wind energy and the Akasofu parameter, epsilon, (which reflects the coupling of solar wind energy into the magnetosphere) were studied. Supersubstorm are associated with increased solar wind speed and thus increased energy levels. The energy levels are much higher for days with supersubstorm than a quiet day. The coupling parameter (epsilon) strongly correlates with the southward magnetic field component (IMF-Bz) indicating that the increased coupling of energy leads to higher IMF-Bz levels. An interesting insight was obtained showing that the increase in solar wind energy level need not necessarily lead to increased coupling of energy into magnetosphere. Solar energy components and geo-magnetic indices sometimes showed a lagged relationship. Thus increase in solar wind energy or coupled energy into magnetosphere sometimes appear as geo-magnetic index changes with a time-lag.

Energy component is the source of disturbances and geo-magnetic index changes brought about during a geo-magnetic storm. Though this thesis did an exploratory qualitative analysis of the energy dynamics and its relation to geo-magnetic indices, and produced insights of the dynamics using correlation, cross-correlation, and summary statistical analysis, further work needs to be done to get more understanding of the supersubstorm. We obtained a negative correlation between the solar wind energy and the epsilon parameter. It has to be further explored why higher solar wind energy level do not lead to increased energy coupling. This can be better understood by studying all the energy components in the solar wind energy. Similarly, we obtained the energy coupling parameter greater than solar wind energy for one of the supersubstorm. We need to investigate if this is due to the inaccurate measurements (as there are few data points) or if there are opportunities to further the main equation to compute the coupling coefficient.

From the cross-correlation analysis of energy and geo-magnetic indices, we found variable lag between different components for different event days (85 minutes to 240 minutes). It is to be further investigated what factors lead to these variable lags and if these are dependent on factors like the energy levels. Other investigations could look into deriving quantitative relationship between the energy components and geo-magnetic indices. Geo-magnetic indices registered at the earth are easier and continuously measured. A quantitative mapping between the indices and the energy components would help to easily understand different solar activity dynamics in terms of energy.

# References

[1] E. Priest, *Magnetic reconnection at the Sun.* Magnetic reconnection in space and laboratory plasmas, **30**: p. 63. (1984).

[2] Y. Kamide and S.I. Akasofu, *Notes on the auroral electrojet indices.* Reviews of Geophysics, **21**(7): p. 1647-1656. (1983)

[3] J.R. Kan, et al., *Magnetospheric substorms.* Washington DC American Geophysical Union Geophysical Monograph Series, **64**. (1991).

[4] S.-I. Akasofu, *Energy coupling between the solar wind and the magnetosphere.* Space Science Reviews, **28**(2): p. 121-190. (1981).

[5] B. Adhikari, P. Baruwal, and N.P. Chapagain, *Analysis of supersubstorm events with reference to polar cap potential and polar cap index.* Earth and Space Science, **4**(1): p. 2-15. (2017).

[6] S. Chapman and V.C. Ferraro, *A new theory of magnetic storms.* Terrestrial Magnetism and Atmospheric Electricity, **36**(2): p. 77-97. ( 1931).

[7] J.W. Dungey, *Interplanetary magnetic field and the auroral zones.* Physical Review Letters, **6**(2): p. 47. (1961).

[8] E. Parker, *Dynamics of the geomagnetic storm.* Space Science Reviews, **1**(1): p. 62-99. (1962).

[9] S.I. Akasofu and S. Chapman, *The development of the main phase of magnetic storms.* Journal of Geophysical Research, **68**(1): p. 125-129. (1963).

[10] S. Akasofu, *Magnetic storms: the simultaneous development of the main phase (DR) and of polar magnetic storm (DP).* J. Geophys. Res., **68**: p. 3158-3185. (1963).

[11] A. Hewish and S. Bravo, *The sources of large-scale heliospheric disturbances.* Solar physics, **106**(1): p. 185-200. (1986).

[12] L. Burlaga, K. Behannon, and L. Klein, *Compound streams, magnetic clouds, and major geomagnetic storms.* Journal of Geophysical Research: Space Physics, **92**(A6): p. 5725-5734. (1987).

[13] W. Gonzalez, et al., *Interplanetary phenomena associated with very intense geomagnetic storms.* Journal of Atmospheric and Solar-Terrestrial Physics, **64**(2): p. 173-181. (2002).

[14] W. Gonzalez, et al., *What is a geomagnetic storm?* Journal of Geophysical Research: Space Physics, **99**(A4): p. 5771-5792. (1994).

[15]  R. Rawat, S. Alex, and G. Lakhina, *Low-latitude geomagnetic response to the interplanetary conditions during very intense magnetic storms.* Advances in Space Research, **43**(10): p. 1575-1587. (2009).

[16]  I. Richardson, E. Cliver, and H. Cane, *Sources of geomagnetic storms for solar minimum and maximum conditions during 1972–2000.* Geophysical Research Letters, **28**(13): p. 2569-2572. (2001).

[17]  R. Lopez, et al., *Solar wind density control of energy transfer to the magnetosphere.* Geophysical research letters, **31**(8). (2004).

[18]  D. Xu, et al., *Statistical relationship between solar wind conditions and geomagnetic storms in 1998–2008.* Planetary and Space Science, **57**(12): p. 1500-1513. (2009).

[19]  Z. Ziming, C. Jinbin, and L. Yi, *Auroral electrojet event associated with magnetospheric substorms.* 空间科学学报, **30**(4): p. 349-355. (2010).

[20]  B. Tsurutani, et al. *Extremely intense (SML≲–2500 nT) substorms: isolated events that are externally triggered?* in *Annales Geophysicae*. Copernicus GmbH. (2015).

[21]  R. Hajra, et al., *Supersubstorms (SML<− 2500 nT): Magnetic storm and solar cycle dependences.* Journal of Geophysical Research: Space Physics, **121**(8): p. 7805-7816. (2016).

[22]  E. Spiegel and J.-P. Zahn, *The solar tachocline.* Astronomy and Astrophysics, **265**: p. 106-114. (1992).

[23]  E. Priest, *Magnetohydrodynamics of the Sun*. Cambridge University Press, (2014).

[24]  S.-I. Akasofu, *The development of the auroral substorm.* Planetary and Space Science, **12**(4): p. 273-282.(1964).

[25]  D. McComas, et al., *Weaker solar wind from the polar coronal holes and the whole Sun.* Geophysical Research Letters, **35**(18). (2008).

[26]  J.L. Kohl, et al., *Ultraviolet spectroscopy of the extended solar corona.* The Astronomy and Astrophysics Review, **13**(1-2): p. 31-157. (2006).

[27]  G.K. Parks, *Physics of space plasmas-an introduction.* Redwood City, CA, Addison-Wesley Publishing Co., (1991), 547 p., (1991).

[28]  E. Priest, *Solar Magnetohydrodynamics, Reidel Publ.* Co., Dordrecht, Holland, 1982.

[29]  E.N. Parker, *Cosmical magnetic fields: Their origin and their activity.* Oxford, Clarendon Press; New York, Oxford University Press, (1979), 858 p., (1979).

[30]  K. Mursula and T. Ulich, *A new method to determine the solar cycle length.* Geophysical Research Letters, **25**(11): p. 1837-1840. (1998).

[31] R.c. Tousey, et al., *A preliminary study of the extreme ultraviolet spectroheliograms from Skylab.* Solar Physics, **33**(2): p. 265-280. (1973).

[32] J. Gosling, et al., *Mass ejections from the Sun: A view from Skylab.* Journal of Geophysical Research, **79**(31): p. 4581-4587. (1974).

[33] J. Zhang, et al., *On the temporal relationship between coronal mass ejections and flares.* The Astrophysical Journal, **559**(1): p. 452. (2001).

[34] R.A. Harrison, *Coronal magnetic storms: A new perspective on flares and the 'solar flare myth' debate.* Solar Physics, **166**(2): p. 441-444. (1996).

[35] D.F. Webb and R.A. Howard, *The solar cycle variation of coronal mass ejections and the solar wind mass flux.* Journal of Geophysical Research: Space Physics, **99**(A3): p. 4201-4220. (1994).

[36] A.S. Jursa, *Handbook of geophysics and the space environment.*: Air Force Geophysics Laboratory, Air Force Systems Command, United States …. Vol. 1. (1985)

[37] P.A. Cassak and S.A. Fuselier, *Reconnection at Earth's Dayside Magnetopause*, in *Magnetic Reconnection: Concepts and Applications*, W. Gonzalez and E. Parker, Editors., Springer International Publishing: Cham. p. 213-276. (2016).

[38] S. Cowley, *Evidence for the occurrence and importance of reconnection between the Earth's magnetic field and the interplanetary magnetic field.* Washington DC American Geophysical Union Geophysical Monograph Series, **30**: p. 375-378. (1984).

[39] W. Hughes, *The magnetopause, magnetotail, and magnetic reconnection.* Introduction to Space Physics,: p. 227-287. (1995).

[40] H. Frey, et al., *Continuous magnetic reconnection at Earth's magnetopause.* Nature, **426**(6966): p. 533. (2003).

[41] H. Frey, et al., *Magnetospheric Physics (SMP)-SMP 2. Proton aurora in the cusp (DOI 10.1029/2001JA900161).* Journal of Geophysical Research-Part A-Space Physics, **107**(7). (2002).

[42] S. Fuselier, et al., *Cusp aurora dependence on interplanetary magnetic field Bz.* Journal of Geophysical Research: Space Physics, **107**(A7): p. SIA 6-1-SIA 6-10. (2002).

[43] B.T. Tsurutani, et al., *Origin of interplanetary southward magnetic fields responsible for major magnetic storms near solar maximum (1978–1979).* Journal of Geophysical Research: Space Physics, **93**(A8): p. 8519-8531. (1988).

[44] T. Araki, *Global structure of geomagnetic sudden commencements.* Planetary and Space Science, **25**(4): p. 373-384. (1977).

[45] G. Rostoker, et al., *Magnetospheric substorms—Definition and signatures.* Journal of Geophysical Research: Space Physics, **85**(A4): p. 1663-1668. (1980).

[46] N. Østgaard, et al., *Observations and model predictions of substorm auroral asymmetries in the conjugate hemispheres.* Geophysical research letters, **32**(5). (2005).

[47] S. Akasofu, *Polar and magnetospheric substorms, Reidel Publ.* Co., Dordrecht, Holland, (1968).

[48] B. Tsurutani and C.I. Meng, *Interplanetary magnetic-field variations and substorm activity.* Journal of Geophysical Research, **77**(16): p. 2964-2970. (1972).

[49] R.H.a.B.T.T.a.E.E.a.W.D.G.a.J.W. Gjerloev, *Supersubstorms (SML < −2500 nT): Magnetic storm and solar cycle dependences.* Journal of Geophysical Research: Space Physics, (2016).

[50] A.K. Gregorczyk, *Ionosphere response to three extreme events occurring near spring equinox in 2012, 2013 and 2015, observed by regional GNSS-TEC model.* Journal of Geodesy, (2018).

[51] E.E. Bruce T. Tsurutani, Kazunari Shibat, Olga P. Verkhoglyadova, Anthony J. Mannucci and Walter D. Gonzalez Janet U. Kozyra and Martin Pätzold, *The interplanetary causes of geomagnetic activity during the 7–17 March 2012 interval: a CAWSES II overview.* Journal of Space Weather Space Climate, (2014).

[52] B.T.T.a.R.H.a.E.E.a.J.W. Gjerloev, *Extremely intense (SML ≤ −2500 nT) substorms: isolated events that are externally triggered?* AnGeo Comm, (2015).

[53] P. Perreault and S. Akasofu, *A study of geomagnetic storms.* Geophysical Journal International, **54**(3): p. 547-573. (1978).

[54] R.M. Mac-Mahon and W. Gonzalez, *Energetics during the main phase of geomagnetic superstorms.* Journal of Geophysical Research: Space Physics, **102**(A7): p. 14199-14207. (1997).